\documentclass[amsmath,amssymb,11pt]{article}
\usepackage{jheppub2}
\pdfoutput=1
\usepackage{graphicx,epsfig}
\usepackage{float}
\usepackage{subfig}
\usepackage{subfloat}
\usepackage[utf8]{inputenc}
\usepackage{hyperref}
\usepackage{caption}
\usepackage[dvipsnames]{xcolor}

\usepackage{lmodern}

\newcommand{\beq}{\begin{equation}}

\newcommand{\eeq}{\end{equation}}

 \newcommand{\be}{\begin{equation}}
 \newcommand{\ee}{\end{equation}}
 \newcommand{\bea}{\begin{eqnarray}}
 \newcommand{\eea}{\end{eqnarray}}

\usepackage{tikz}
\usetikzlibrary{positioning}
\usetikzlibrary{intersections}
\usetikzlibrary{fadings} 
\usetikzlibrary{arrows.meta} 
\usetikzlibrary{arrows}

\tikzfading[name=fade out,
inner color=transparent!0,
outer color=transparent!100]

\definecolor{cherryblossompink}{rgb}{1.0, 0.72, 0.77}
\definecolor{lightblue}{rgb}{0.68, 0.85, 0.9}

\usetikzlibrary{decorations.pathmorphing}
\usetikzlibrary{decorations.pathreplacing,decorations.markings}

\usetikzlibrary{backgrounds,automata}

\title{Emergence of phantom cold dark matter from spacetime diffusion}

\author{Jonathan Oppenheim,}
\emailAdd{j.oppenheim@ucl.ac.uk}
\author{Emanuele Panella,}
\emailAdd{emanuele.panella.21@ucl.ac.uk}
\author{and Andrew Pontzen}
\emailAdd{a.pontzen@ucl.ac.uk}

\affiliation{
Department of Physics and Astronomy, University College London,\\
Gower Street, London, WC1E 6BT, United Kingdom}

\abstract{General relativity can be reconciled with quantum field theory without quantising the geometry only if the metric evolves stochastically. In this article, we explore the consequences of such a proposal at early cosmological times. We find the stochastic evolution results in the spatial metric diffusing away from its deterministic value, generating phantom cold dark matter (CDM). It is produced primarily at the end of the inflationary phase of the Universe's evolution, with a statistical distribution that depends on the specifics of the early-times cosmological model.
We find the energy density of this phantom cold dark matter is positive on average, a necessary condition to reproduce the cosmological phenomenology of CDM, although further work is required to calculate its mean density and spatial distribution. If the density is cosmologically significant, phantom dark matter acts on the geometry in a way that is indistinguishable from conventional CDM. As such, it has the potential to reproduce phenomenology such as structure formation, lensing, and galactic rotation curves. We conclude by discussing the possibility of testing hybrid theories of gravity by combining measurements of the cosmic microwave background with tabletop experiments.

}

\begin{document}

\maketitle

\section{Introduction} \label{sec:intro}

General relativity provides a dynamic background geometry in which matter fields evolve, determining the causal structure essential to present formulations of quantum field theory. Taking the geometric nature of gravity as fundamental, it is not certain that spacetime geometry must be quantised along with the rest of the Standard Model of particle physics. In fact, a fundamentally classical theory of gravity has appeals of its own~\cite{oppenheim2023is-it-time}, possibly addressing gaps in our understanding of quantum mechanics, with the measurement problem being the most evident~\cite{singh2015possible}. Further, a theory in which a classical metric consistently interacts with quantum matter manifestly avoids the {\it problem of time}~\cite{isham1992canonical,kuchar1992time,anderson2012problem} altogether, which arises when the Hamiltonian constraint is imposed quantum mechanically via the Wheeler-DeWitt equation. In fact, as we argue in this article, violation of the standard constraints of general relativity is a natural consequence of a hybrid theory of gravity, leading to interesting phenomenology.

Standard arguments in favour of the need to quantise the gravitational field~\cite{eppley1977tnecessity,bergmann1957summary} can be circumvented, enabling consistent models of classical gravity interacting with quantum matter to be constructed~\cite{diosi1995quantum,alicki2003completely,kafri2014classical,tilloy2016sourcing,tilloy2017principle}. Not only can a framework of classical-quantum interactions be made consistent~\cite{blanchard1995event-enhanced,diosi2014hybrid}, but the form of hybrid dynamics is highly constrained under some physically motivated assumptions~\cite{oppenheim2021postquantum,oppenheim2022classes}. A crucial prediction of the resulting classical-quantum (CQ) theories is that the classical degrees of freedom need to evolve stochastically, with the amount of noise indissolubly related to the typical rate of decoherence of the quantum system to which it is coupled~\cite{diosi1995quantum,oppenheim2021postquantum,oppenheim2023gravitationally}. A version of this theory applied to general relativity, known as ``postquantum classical gravity''~\cite{oppenheim2022classes,oppenheim2023path,oppenheim2023covariant,oppenheim2024diffeomorphism,oppenheim2022constraints,layton2023weak}, or, alternatively, classical-quantum (CQ) gravity, is the natural choice for the null hypothesis against which quantum gravity should be tested. Indeed, although the form of the dynamics can also provide an effective theory for two quantum systems interacting with each other in the classical limit of one of the two~\cite{layton2024classical}, the relation between the free parameters of the theory known as the “decoherence-diffusion trade-off”~\cite{oppenheim2023gravitationally} needs to hold only if the classical system is fundamentally so. Finding a violation of the latter, would indirectly support the quantumness of gravity. The pure gravity theory of~\cite{oppenheim2023covariant} was also shown to be renormalisable in $3+1$ dimensions~\cite{grudka2024renormalisation}, in contrast to perturbative quantum gravity~\cite{goroff1985ultraviolet}. 

As mentioned, recent work proved that there exists a regime where this framework is also the correct effective description of a fully quantum system when a part of it is treated classically. Semiclassical gravity is usually studied through the semiclassical Einstein's equations, which are famously valid only when quantum fluctuations are small~\cite{kuo1993semiclassical,ahmed2024semiclassical} and can be exactly solved only in a handful of cases (e.g.~\cite{christensen1977trace,emparan2002quantum,emparan2020quantum,emparan2022black,panella2023quantum,climent2024chemical,panella2024three-dimensional}). CQ dynamics could provide insights as an effective theory even if gravity is indeed quantum, extending the validity of semiclassical gravity beyond the restricted regime where the semiclassical Einstein's equations apply. 

Earlier works explored the effects of stochastic fluctuations on classical spacetime, showing that they can induce an effective cosmological constant and allow for modified gravitational potentials which affect galactic rotation curves in a way that mimics the effect of dark matter~\cite{oppenheim2024anomalous}. However, since cold dark matter's effects are not restricted to galactic scales, but leave imprints also on lensing, structure formation and cosmic microwave background (CMB) statistics, it is important to explore whether theories akin to~\cite{oppenheim2021postquantum,oppenheim2023covariant} can satisfy such cosmological constraints.

In this article, we study the stochastic Friedmann–Lema\^itre–Robertson–Walker (FLRW) Universe, probing the cosmological consequences of such stochasticity when modelled as a source in Einstein's field equations. Other proposals in which the metric field is coupled to a stochastic source have been put forward over the years, such as two different models named “stochastic gravity" ~\cite{hu2008stochastic,moffat1997stochastic}, the everpresent Lambda~\cite{ahmed2004everpresent} and models motivated by unimodular gravity~\cite{perez2021resolving,landau2022cosmological}. We diverge from those approaches both in interpretation and implementation, due to the hypothesis of the gravitational degrees of freedom being \textit{fundamentally} classical. This  strongly constrains the form of the moments of the probability distribution of the noise field. Our starting point is that the dynamics needs to be completely positive and norm preserving, and linear in the probability density to preserve the statistical interpretation of the probability density. This forces the dynamics to be of the form of~\cite{oppenheim2021postquantum,oppenheim2022classes}, where the statistics of the noise field cannot depend on the quantum states. We will compare some of the stochastic gravity models which have been proposed over the years with ours in Section \ref{sec:discussion}.

In standard GR, the set of algebraic relations known as the \textit{constraint equations}, which the gravitational state has to satisfy at all times, arise as a consistency condition for diffeomorphism invariance. Further, the classical equations of motion guarantee that if the system is initialised on the constraint surface, it remains on it. This is no longer necessarily true in a stochastic theory. The generators of the diffeomorphisms become stochastic, and the GR constraint can be violated without necessarily breaking diffeomorphism invariance~\cite{oppenheim2022constraints}, a point we shall return to in Section \ref{sec:discussion}. 

In our model, the equations of motion evolve the Hamiltonian constraint $C_H$ off the $C_H=0$ surface even on average, with a positive drift term. Violation of the constraint can behave in an identical manner to cold (i.e. pressureless) dark matter, an effect previously discussed (with a different motivation) by ~\cite{burns2023time,kaplan2023classical} and further explored in~\cite{casadio2024relaxation,delgrosso2024cosmological}. 
In general however, the constraint violations can lead to the appearance of both positive and negative energy density. The main result of the present paper is to show that cosmological diffusion provides a natural mechanism to drive the system off the constraint surface {\it positively on average}, a necessary condition for the constraint violation to appear as if it were cold dark matter.

Following from this, we find several other results. Starting from plausible assumptions about the dependence of diffusion rate on horizon scale, we calculate the amount of phantom cold dark matter that is produced. We find that the resulting density depends only on the dimensionless coupling constant of the theory, and the number of e-folds during radiation domination, with the phantom cold dark matter being produced primarily at the end of the inflationary phase. Next, we find that shortly after inflation production halts, a necessary condition for reproducing the CMB power spectrum. We then highlight how the combination of tabletop experiments and cosmological data can provide non-trivial tests for the model we discuss. However, there are still a number of theoretical issues that need to be resolved before such an analysis can be reliably performed. We therefore conclude with a summary of open questions and a discussion of covariance of stochastic theories.

Unless explicitly stated otherwise, we take $c=1$ and $(-,+,+,+)$ as the metric signature.

\section{Background} \label{sec:theory}
\subsection{Stochastization of Einstein's equations} 

The dynamics of any theory that describes a classical metric interacting with quantum matter without quantising the gravitational theory must be irreversible as shown in~\cite{oppenheim2021postquantum,oppenheim2023gravitationally} (see~\cite{galley2023consistent} for an alternative argument). In particular, the coupling between quantum and classical degrees of freedom requires stochasticity in order to support coherent quantum superpositions in the basis of the interaction. In this work, we do not need the full formalism first presented in~\cite{oppenheim2021postquantum,oppenheim2022classes} (see~\cite{layton2023weak,oppenheim2024diffeomorphism,grudka2024renormalisation,oppenheim2024anomalous,wellerdavies2024quantum} for recent applications to gravity), as we assume quantum degrees of freedom to be fully decohered on cosmological scales. Rather, we look at a stochastic modification of Einstein's field equations (EFEs), motivated by the results in~\cite{layton2023healthier} which show that the irreversibility of continuous and Markovian hybrid dynamics translates into an equation of motion for the classical system sourced by a background white noise field $\xi$, irrespective of the state of the quantum system. We wish to consider the noise generated by a Gaussian stochastic scalar that satisfies:
\begin{equation}
\begin{split}
\label{eq:stats}
    &\mathbb{E}[\xi(x,t)]=0 \ , \\ 
    &\mathbb{E}[\xi(x,t)\xi(x',t')]= \frac{D_2}{\sqrt{h} N} D(x,t,x',t') \ ,
\end{split}
\end{equation}
where the expectation value $\mathbb{E}[\cdot]$ is to be considered across realisations of the noise field. Here, $D_2$ is the diffusion coefficient, and $D$ is a function which encodes the correlation in the noise tensor between spacetime points. $N$ and $h$, the lapse function and the determinant of the spatial metric, will be defined shortly, but their appearance is required for covariance. To preserve covariance, we also adopt the local kernel:
\begin{equation}
    D(x,t,x',t')=\delta(x-x',t-t')
\end{equation}
as the natural choice. We give an explicit example of how $\xi$ can be constructed using standard tools from stochastic calculus in Appendix \ref{app:noise}.

We work in the Arnowitt–Deser–Misner (ADM) formalism~\cite{arnowitt1959dynamical}, a canonical description of GR. We explicitly pick a foliation of spacetime, inducing the following 3+1 decomposition of the metric:
\begin{equation}
    g_{00}=-N^2+h^{ij}N_i N_j \ , \qquad g_{0i}=N_i \ , \qquad g_{ij}= h_{ij} .
\end{equation}
Here, $h_{ij}$ is the spatial metric on the chosen foliation, while $N$ and $N^i$ (called the lapse function and shift vector respectively) tell us how the three geometry is embedded in the 4-dimensional manifold. The canonical equations of motion for gravity, with minimal coupling to the noise process, are given by:
\begin{align}
    \dot{h}_{ij} &= \{h_{ij}, H_{\text{GR}}\}+ \{h_{ij}, H_{\text{m}} \}  \   ,\\
    \dot{\pi}^{ij} &= \{\pi^{ij}, H_{\text{GR}}\}+ \{\pi^{ij}, H_{\text{m}}\} +  \frac{1}{8\pi G}N\sqrt{h}h^{ij} \xi\ \label{eq:pi_ij}, 
\end{align} 
where $\pi_{ij}$ is the conjugate momentum of the 3-metric and an overhead dot means differentiation with respect to coordinate time $t$. Moreover, $H_{\text{GR}}$ and $H_{\text{m}}$ are the gravity and matter Hamiltonian respectively, with $\{\cdot,\cdot\}$ being the Poisson brackets. We do not add a stochastic term to the evolution equation for $h_{ij}$ in order to yield a geometry that is differentiable. For consistency with other work, we adapt the convention in which the local noise field $\xi$ carries units of inverse area, and $D_2$ is dimensionless.

The third term in \eqref{eq:pi_ij} is the simplest stochastic term we can add that has the right transformation properties (i.e. it is a spatial 2-tensor density with weight $-1$ which also transforms as a scalar density of weight $-1$ under time reparametrisations). Naturally, if $\xi=0$ we recover the standard ADM equations of motion. The same modification can be obtained starting from the gravitational theory with action $S=S_{EH}+S_M$, made up of an Einstein-Hilbert gravitational term and matter term respectively, by adding a term that minimally couples the stochastic field to the metric:
\begin{equation}
\label{eq:stoc_act}
    S_\text{N} = \frac{1}{8\pi G} \int \text{d}^4x \sqrt{g} \xi \ .
\end{equation}
Then, deriving the equations of motion as usual, one arrives at $\eqref{eq:pi_ij}$. Integrals such as \eqref{eq:stoc_act} require careful treatment, since $\xi$ is nowhere integrable in the standard sense. In this article we adopt the It\^{o} definition of stochastic integrals (and, consequently, stochastic differential equations), since it is the natural one when interpreting the noise process as fundamental. Further detail is given in Appendix \ref{app:ito}.

Due to covariance, in standard GR the equations of motion are supplemented by the vanishing of the Hamiltonian and momentum constraints, $C_H$ and $C_P^i$ respectively:
\begin{align}
    C_{H} &\equiv\mathcal{H} + N^2 \sqrt{h}T^{00} \approx 0 \ , \\
    C_{P}^i &\equiv \mathcal{H}^i - N \sqrt{h} h^{ij}T_{0j} \approx 0 \ ,
\end{align}
where $\approx$ indicates that such algebraic relations hold only on-shell. Here, $\mathcal{H}$ and $\mathcal{H}^i$ are functionals of the gravitational phase space variables $h_{ij}$ and $\pi_{ij}$ only. The role of constraints in CQ theory has been explored in detail in~\cite{oppenheim2022constraints}, and we will further explore the relation between constraints and covariance in stochastic theories in Section \ref{sec:discussion}. 

\subsection{Stochastic FLRW}\label{sec:FLRW}
We now focus on Friedmann-Lema\^{i}tre-Robertson-Walker universes. As usual, we impose spatial isotropy and homogeneity, picking the gauge $N^i=0$ to have these symmetries manifest in the 3-metric. The spatial metric $h_{ij}$ must then be of the form:
\begin{equation}
\begin{split}
    \text{d}\ell^2 &= h_{ij} \text{d}x^i \text{d}x^j \\
    &= a^2(t) \left(\frac{1}{1-kr^2} \text{d}r^2 + r^2 \text{d}\Omega^2\right) \ ,
\end{split}
\end{equation}
where $\text{d}\Omega^2$ is the metric on $S_2$, whilst $a(t)$ is the so-called scale factor, a dimensionless function controlling the time-dependent “radius” of the Universe. On the other hand, $k$ is the Gaussian curvature of the space for $a(t)=1$, and, in our convention, has units of inverse area. Later, we consider early time cosmology with an inflationary phase. Since inflation washes out any spatial curvature, we will later restrict our attention to flat spatial slices ($k=0$) for simplicity, but will treat the general case for as long as possible. Homogeneity also requires that the lapse $N$ has no spatial dependence. Up to the choice of lapse function, the 4-metric has then the form:
\begin{equation}
    \text{d}s^2 = -N(t)^2 \text{d}t^2 + a^2(t) \left(\frac{1}{1-kr^2} \text{d}r^2 + r^2 \text{d}\Omega^2\right) \ .
\end{equation}
Note that the lapse function $N$ here is also taken to be dimensionless.

For comoving pressureless dust, the stress tensor is given by:
\begin{equation}
\begin{split}
    T_{\mu \nu}&= (1+w) \rho  \ u_{\mu} u_{\nu} +w \rho \ g_{\mu \nu} \\
    &=N^2(t) \rho \ \delta_{\mu 0}\delta_{\nu 0} + w\rho \ g_{\mu\nu} ,
\end{split}
\end{equation}
where we have used the fact that the appropriately normalised 4-velocity $u^\mu$ of a comoving fluid is given by:
\begin{equation}
    u^{\mu}=\frac{1}{N} (1,0,0,0) \ .
\end{equation}
By plugging in the FLRW metric in the Einstein-Hilbert action, one can show that the mini-superspace Hamiltonian is given by (allowing a cosmological constant $\Lambda$ in units of inverse area):
\begin{equation}
     H_\text{GR}=-N \left( \frac{2 \pi G}{3} \frac{\pi_a^2}{a}+ \frac{3ka}{8\pi G}-\frac{\Lambda a^3}{8\pi G} \right) \ .
\end{equation}
Here, we have used the definition of $\pi_a$ as the conjugate momentum of $a$:
\begin{equation}
    \pi_a = \frac{\partial \mathcal{L}_{\text{EH}}}{\partial\dot{a}} = -\frac{3}{4\pi G} \frac{\dot{a}a}{N} ,
\end{equation}
which is related to the conjugate momentum of the homogeneous 3-metric itself via $\pi^{ij}=\delta^{ij} \pi_a/6a $. Now, taking the trace of the ADM equations of motion in the presence of matter with energy $\rho$ and equations of state parameter $w$, we obtain the cosmological evolution equations:
\begin{align}
\label{eq:eom}
    \dot{a} &= -\frac{4\pi G}{3} N \frac{\pi_a}{a} \\
    \dot{\pi}_a &= - \frac{2 \pi G}{3} N\frac{\pi_a^2}{a^2}+N 3 wa^2 \rho + N \frac{3}{8\pi G} k - N\frac{3}{8\pi G}a^2\Lambda+N \frac{3}{8\pi G} a^2 \bar{\xi} \ ,
\end{align}
up to numerical prefactors in the coupling to the noise which can always be absorbed into the diffusion coefficient. Here, we have also included the effect of curvature for completeness. Moreover, we have forced the noise process -- with units of inverse area -- to be homogeneous in space. 

The global random field $\bar{\xi}$ has to obey the following statistics:
\begin{equation}
\label{eq:noise_gen}
     \mathbb{E}[\bar{\xi}(t),\bar{\xi}(t')] = \frac{\bar{D}_2 (a)}{N} \delta(t-t') \ ,
\end{equation}
where the renormalised diffusion coefficient $\bar{D}_2(a)$ is, in principle, an arbitrary functional of the scale factor which encodes how one translates the local theory into the homogeneous one, as we explore in Section \ref{sec:renormalisation}. The factor of inverse lapse is needed for time reparametrisation invariance as discussed in Appendix \ref{app:noise}.

In GR, the Hamiltonian constraint provides an initial condition for the state:
\begin{equation}
\label{eq:con}
    C_H= \frac{2 \pi G}{3} \frac{\pi_a^2}{a}-\rho a^3 + \frac{3ka}{8\pi G}  - \frac{\Lambda a^3}{8\pi G}  \approx 0\ ,
\end{equation}
where we have made the effect of the curvature $k$ explicit.
In this article, we discuss the consequences of a dynamical violation of the constraint. It is, therefore, natural to question whether one is allowed at all to use the constraint as an initial condition. In fact, we show that inflation washes out any initial deviation from the constraint, making the ambiguity in the initial state essentially irrelevant.
A point that will be important later is that comoving pressureless dust enters the equations of the system only through the Hamiltonian constraint. The set of relations given by \eqref{eq:eom} (without noise) and \eqref{eq:con} is completely equivalent to Friedmann's equations.

Since we do not couple matter and the stochastic field directly, we leave the equations of motion for dust unchanged with respect to the standard treatment. Fundamentally, one can understand the system as being described by the following action:
\begin{equation}
    S=S_{\text{EH}}+S_{\text{N}}+S_\text{BK} \ ,
\end{equation}
where $S_\text{BK}$ is the Brown-Kucha\v{r} action~\cite{brown1995dust}, which provides the Lagrangian formulation of a perfect fluid, whilst $S_{\text{EH}}$ is the Einstein-Hilbert action and $S_{\text{N}}$ is the stochastic term defined by Eq.~\eqref{eq:stoc_act}. Consequently, one finds that covariant conservation of the stress-energy tensor associated with the fluid ($\nabla_\mu T^{\mu}_\nu = 0$) still holds. Therefore, the energy density of the fluid dilutes with the scale factor as usual:
\begin{equation}
    \rho = \rho_0 a^{-3(1+w)} \ ,
\end{equation}
with $\rho_0$ being the value of the energy density for unit scale factor. The stochastic field pumps energy only into the gravitational sector. 

\subsection{Renormalising the diffusion coefficient}
\label{sec:renormalisation}
In order to establish the relation between the local diffusion coefficient and the global parameter $\bar{D}_2 (a)$ a procedure to flow to the long-wavelength regime is needed. At present, we do not have a rigorous procedure for performing this renormalisation. Differing procedures could lead to a different scaling of the diffusion with the scale factor. Therefore, we will treat $D_2$ as a general function of $a$ for as long as possible.

Nonetheless we can commute the spatial averaging and time evolution for a well-motivated estimate of the background evolution of the universe (as is common in standard cosmological calculations). Just as the mean energy density enters the equations of motion for a FLRW Universe, we take the average value of the noise for its realisation at time $t$ to be the stochastic source in the global Einstein's equations. Therefore, we interpret $\bar{\xi}$ as the average of the local random field given by \eqref{eq:stats} over a spatial domain on the spacelike constant $t$ hypersurface. We adopt the standard assumption in cosmology that the inhomogeneous Universe can be approximated as homogeneous and isotropic to leading order~\cite{maartens2011is-the-universe}. This should be an acceptable approximation as long as the typical size of the local fluctuations is much smaller than the average energy density of the matter sourcing the homogeneous evolution. Additional work will be required to address whether it is possible to derive formally that such
an effective dynamics is valid in the IR (long-wavelength) limit.

Working in the separate-universe approximation~\cite{wands2000approach}, we imagine the scale factor as assigned to a finite-sized patch of the Universe. To turn the noise field homogeneous, we define:
\begin{equation}
    \bar{\xi}_\Sigma (t) = \frac{\int_{\Sigma} d^3x \sqrt{g} \xi}{\int_\Sigma d^3x \sqrt{g}} \ ,
\end{equation}
where $\Sigma$ is the spatial region over which we average the noise, introducing a long-distance IR cutoff $R_\text{IR}$ over which we trust the homogeneous description. Trivially we still have $\mathbb{E}[\xi_\Sigma (t)]=0$, whereas:
\begin{equation}
    \mathbb{E}[\bar{\xi}_\Sigma(t),\bar{\xi}_\Sigma(t')] =\frac{D_2}{N \int_\Sigma d^3 x \sqrt{g(x)}} \delta(t-t') \ .
\end{equation}
We thus see that when flowing to a model in which local noise becomes averaged noise, $D_2$ needs to be renormalised by the volume of the region over which we average.
The renormalisation scheme then boils down to choosing the spatial domain over which to average; we now present two natural options that lead to very different late-time behaviour.

One natural choice for $R_\text{IR}$ in FLRW cosmologies is the comoving Hubble radius $R_\text{IR} = 1/aH$, i.e. averaging over the spatial region that is in causal contact over the current $e$-fold of cosmological evolution. This leads to an effective variance that scales as the inverse of the Hubble volume:
\begin{equation}
    \label{eq:noise_hub}
     \mathbb{E}[\bar{\xi}_\Sigma(t),\bar{\xi}_\Sigma(t')] = \frac{3}{4\pi}\frac{D_2}{N} H^3 \delta(t-t') \ .
\end{equation}
Imagine now that we work in the low-noise limit, meaning that the stochastic trajectories are well approximated to leading order by the deterministic evolution. In an inflationary early-Universe phase the Hubble parameter is constant ($H(t)=H_I$) and so the variance, too, remains constant. During matter and radiation domination, however, the situation changes radically as the Hubble parameter falls with time following $H\propto t^{-1}$ in both eras. Consequently, the effective noise gets damped significantly as the Universe expands. This is, effectively, a consequence of the causal horizon of the patch expanding after inflation. Indeed, more and more modes re-enter the horizon and contribute to the effective cosmological noise, which converges to the average value with vanishing variance by the central limit theorem. Of course, this is only true at the cosmological level, meaning that the inhomogeneous perturbations still follow a stochastic evolution.

An alternative spatial averaging scheme in which the stochastic term remains relevant at late times is to adopt the choice of $R_{\text{IR}}$ being some fixed comoving radius $R_{\text{IR}} = R/a$. This way, we obtain:
\begin{equation}
     \mathbb{E}[\bar{\xi}_\Sigma(t),\bar{\xi}_\Sigma(t')] = \frac{3}{4\pi R^3}\frac{D_2}{N} \delta(t-t') \ .
\end{equation}
This appears less motivated than the previous choice of averaging over a horizon volume, but we include it as a possibility in the absence of a fully principled approach at present. From now on, we drop the subscript $\Sigma$ from the noise field.

\section{Phantom CDM from constraint violation}
\label{sec:constraint}
\subsection{Violation of the deterministic Hamiltonian constraint}
\label{sec:violation_CH}
In standard GR, the constraint equations are satisfied at all times. However, using It\^{o}'s lemma (the chain rule for stochastic processes), we can see how the constraint evolves on-shell once noise is introduced (see Appendix \ref{app:constraint} for a step-by-step calculation):
\begin{equation}
    \dot{C}_H= D_2(a) \frac{3}{32\pi G}  a^3 N + \frac{1}{2} \pi_a a N\bar{\xi} \ .
    \label{eq:C-preservation}
\end{equation}
The Hamiltonian constraint $C_H\approx0$ is therefore broken by the stochastic dynamics. Previous work has considered the role of constraints in classical-quantum theories~\cite{oppenheim2021constraints,oppenheim2022constraints}, and whether the constraints of the deterministic part of the dynamics can be violated without breaking covariance. For now, we focus on the implications of such a result, but we do comment further on it in Section \ref{sec:discussion}. The second term causes diffusion of $C_H$ around zero, as expected. The first term is an anomalous drift that can be understood via It\^{o}'s lemma. This “second order force” pushes the average value of the constraint away from zero, meaning that $C_H$ is not conserved even on average. Equation \eqref{eq:C-preservation} is the first result of this article. From it, we can immediately see that the constraint violation will, on average, be positive.

The positive constraint violation is connected with entropy production in the state, since the average constraint depends on the variance of the conjugate momentum. The probability distribution of momenta begins sharply peaked by assumption in the initial conditions, but diffuses in time through stochastic kicks which increase the entropy of the classical distribution. This is formally equivalent to the heating up of a Brownian particle without friction, and it is fundamentally connected to the irreversibility of the Fokker-Planck equation governing the evolution of the probability distribution over states. 

\subsection{Phantom CDM}
Now we consider what effects departing from the Hamiltonian constraint has upon the observable properties of the Universe. Consider first a matter dominated Universe for simplicity. When averaging over the Hubble horizon, the noise term drops out at late times from the equations of motion and the system reduces to standard GR, since $H \to 0$. During the early phases of the cosmological evolution, however, the system might have accumulated a non-trivial violation of the deterministic constraint $\delta C$. 

In such a low-noise regime, a violation of the Hamiltonian constraint evolves as “phantom" extra pressureless dust in the system. To see this, first note that the deterministic Friedmann equations preserve the value of the constraint even when off-shell (Appendix \ref{app:constraint}); consequently, the constraint value $C_H$ is frozen in time once the noise becomes subdominant. The constraint equation reads:
\begin{equation}
    \frac{2\pi G}{3}\frac{\pi_a^2}{a}-\rho a^3 = C_H,
\end{equation}
where $\rho$ is the total matter density, including dark energy. 
An observer who infers the expansion history in such a universe assuming that it is governed by GR would attempt to absorb the non-zero value of $C_H$ into this total matter density. For a multi-component fluid, $\rho$ can be re-written as
\begin{equation}
    \rho a^3 = \sum_i \rho_i a^{-3w_i} \ ,
\end{equation}
meaning that $C_H$ can be absorbed into an effective $w_i=0$ component. Therefore, the state with violation of the constraint $C_H$ corresponds to a standard FLRW geometry with effective dust energy density $\rho_\text{eff,0}=\rho_{\text{m},0}+C_H$, where $\rho_{\text{m},0}$ is the energy density of the physical dust when the scale factor $a$ is unity.

A similar observation has been made in~\cite{burns2023time,kaplan2023classical} and studied in detail in~\cite{casadio2024relaxation,delgrosso2024cosmological}, motivated by  the classical limit of quantum gauge theories. In~\cite{kaplan2023classical} a simple argument is presented for why constraint violations appear like dust, even in inhomogeneous spacetimes. To recap the argument, consider Einstein's equations in covariant form:
\begin{equation}
\label{eq:cov_eq}
    G^{\mu}_\nu = 8\pi G \Tilde{T}^{\mu}_\nu \ ,
\end{equation}
where the $(i,j)$ sector corresponds to the ADM equations of motion, whilst the $(0,\mu)$ is related to constraints. Since we are considering the small-noise regime, the spatial part of Einstein's equations is to be satisfied exactly, whilst we allow for the constraints to be violated. This corresponds to having:
\begin{equation}
    G^i_j = 8\pi G T^{i}_{ j} \ , 
\end{equation}
but
\begin{equation}
    G^{0}_\mu = 8\pi G T^{0}_{\mu} + 8\pi G C^{ 0}_{\mu}  \ , 
\end{equation}
where $T^{\mu}_{\nu}$ is the visible matter stress tensor, whilst $C^\mu_\nu$ is the constraint violation. This is the situation we will encounter soon after inflation, when the state has diffused away from the $C_H\approx 0$ surface on average in the positive direction, while $C_i\approx 0$ is still guaranteed from homogeneity and isotropy. For renormalisations scheme that lead to a $H^n$ scaling for the diffusion coefficient, the post-inflationary stochastic dynamics is suppressed due to averaging the fluctuations over an increasingly larger spatial volume whenever $n\geq 3$, as we show shortly. This includes the case we focus on, namely the one where the averaging happens over the Hubble horizon. 

A strong constraint on how $C^\mu_\nu$ varies in spacetime now arises from the combination of the Bianchi identities and the conservation of stress-energy of the true fluid source. Indeed, by looking at the LHS of \eqref{eq:cov_eq}, we have
\begin{equation}
    \nabla_\nu G^{\nu}_\mu = 0 \implies \nabla_\nu(T^\nu_\mu + C^\nu_\mu) = 0 \ .
\end{equation}
Imposing covariant conservation of the matter stress tensor, we end up with the requirement:
\begin{equation}
    \nabla_\mu C^\mu_\nu = 0 \,
\end{equation}
which, alongside $C^i_j=0$, implies that the constraint violations evolve exactly like a matter perturbation with $w=0$, even in highly inhomogeneous limits. For completeness, we show in Appendix \ref{app:inhomogeneous} that the same exact statement can be derived by using the spatial components of Einstein's equations and covariant conservation of the visible matter in linear perturbation theory, although the argument given above is more general.

\subsection{Production during inflation}
In order for the constraint violation to play the role of CDM, $C_H$ needs to be on average positive and in agreement with the dark matter density inferred from observations. We find that an inflationary phase of the early-Universe can drive the Universe to the desired state, albeit with a density that is highly uncertain. To see why, we return to the homogeneous model in the separate Universe approximation. Consider the ADM equations of motion for a FLRW Universe with Lambda domination instead (which is a good approximation to the inflationary state):
\begin{align}
\label{eq:inflation}
    \dot{a} &= -\frac{4\pi G}{3} \frac{\pi_a}{a} \\
    \dot{\pi}_a &= - \frac{2 \pi G}{3} \frac{\pi_a^2}{a^2}-\frac{3}{8\pi G} a^2\Lambda_I + \frac{3}{8\pi G} a^2 \bar{\xi}  \ ,
\end{align}
where it is now evident that the stochastic field acts as a random fluctuation to the dark energy term. Note that we have chosen $\Lambda_I$ to have the geometric units of $\text{m}^{-2}$ and picked the renormalisation scheme for $D_2$ where we average over the Hubble horizon:
\begin{equation}
    \bar{D}_2(a) =\frac{3}{4\pi} D_2 H^3.
\end{equation}
We can further rescale $\bar{\xi}$ to make the dependence of the variance explicit:
\begin{equation}
    \bar{\xi} = \sqrt{\frac{3}{4\pi}D_2 H^3} \bar{\zeta}
\end{equation}
This way, $\bar{\zeta}$ has units of $\text{m}^{-1/2}$ and moments:
\begin{equation}
     \mathbb{E}[\bar{\zeta}(t)] = 0 \ , \qquad \mathbb{E}[\bar{\zeta}(t),\bar{\zeta}(t')] = \frac{1}{N} \delta(t-t') \ ,
\end{equation}
where the diffusion coefficient is unitless (the Dirac delta carries units of inverse length).

Solving the stochastic differential equation exactly is not feasible. Assuming small diffusion, however, we can approximate the inflationary evolution as the deterministic exponential expansion:
\begin{equation}
    a(t) = \tilde{a} e^{H_I t} \ ,
\end{equation}
as we are interested in the leading order in $D_2$ contribution to the constraint generation. $\tilde{a}$ is here the initial value of the scale factor at inflation and $H_I = \sqrt{\Lambda_I/3}$ the value of the Hubble parameter (which, we have assumed, remains constant during the inflationary period -- a good approximation since the $H(t)=H_I$ solution is an attractor for the deterministic equation). More in detail, to see why this approximation is robust, consider the following change of variable: $\pi_a \to - 3a^2H/4\pi G$, which leads to:
\begin{align}
    \dot{a} &= aH \\
    \dot{H} &= \frac{\Lambda-3H^2}{2}-\sqrt{\frac{3 D_2}{16 \pi}H^3} \bar{\xi}  \ .
\end{align}
The SDE for the Hubble parameter $H$ has been studied in detail in~\cite{bishop2019onedimensional}, where it was shown that, for our parameters, the system reaches equilibrium in $t\approx2/3H_I$. In its steady-state, $H^2$ is Inverse-Gamma distributed, i.e. $H^2\sim\textrm{Inv-Gamma}(1+8\pi/D_2,8\pi \Lambda/3D_2)$. The distribution of $H$ has therefore fat tails, but both mean and the variance are well-behaved:
\begin{equation}
    \mathbb{E}_{s}[H] = \sqrt{\frac{8\pi}{D_2}}H_I \frac{\Gamma\left(\frac{8\pi}{D2}+\frac{1}{2}\right)}{\Gamma\left(\frac{8\pi}{D2}+1\right)}=H_I \left(1-\frac{D_2}{32\pi}\right)+ \mathcal{O}(D_2^2) \ , \qquad \mathbb{E}_{s}[H^2] = H_I^2 \ , 
\end{equation}
where the subscript indicates that these are the moments of the steady-state distribution. Therefore, the trajectories for $H$ are distributed, to leading order in $D_2$, as
\begin{equation}
    \frac{\sqrt{\mathrm{Var}(H)}}{\mathbb{E}_s[H]} \approx \sqrt{\frac{D_2}{32\pi}} \ll 1 \ .
\end{equation}
Then, it follows
\begin{equation}
    \ln a = \int_0^t dt' H(t') \approx \mathbb{E}_s[H] t \approx H_I t \left(1-\frac{D_2}{32\pi}\right) \ ,
\end{equation}
where the replacement with the expectation value is exact in the $t\to\infty$ limit (i.e. $1/t \int dt' H$ is an estimator for the mean), but it is a very good approximation when $H_I t\gg1$ nonetheless.

Then, the evolution  of the constraint is given by:
\begin{equation}
\begin{split}
    \dot{C}_H &= \frac{9 D_2}{128\pi^2 G} a^3 H^3 + \frac{1}{2} \pi_a a \sqrt{\frac{3}{4\pi}D_2 H^3} \bar{\zeta}  \\ 
    &= \frac{9 D_2 }{128\pi^2 G}\tilde{a}^3 e^{3H_I t} H_I^3 -  \frac{3}{8\pi G}H_I \tilde{a}^3 e^{3H_I t} \sqrt{\frac{3}{4\pi}D_2 H_I^3} \bar{\zeta} \ .
\end{split}
\end{equation}
We take inflation to last from $t_0=0$ to reheating at $t_I$, corresponding to $N_I = \ln[a(t_I)/a(0)]$ $e$-folds of expansion. The evolution of the expectation value of $C_H$ from the beginning of inflation can be readily integrated (recalling that $\bar{\zeta}$ has zero mean) to give (to leading order in $D_2$):
\begin{equation}
\begin{split}
    \bar{C}_I = \mathbb{E}[C_H(t_I) ] &= C_{H,0} +  \frac{9 D_2 }{128\pi^2 G}\tilde{a}^3 H_I^3 \int_{0}^{t_I} e^{3H_I t'} \text{d}t' + \mathcal{O}(D_2^2)\\
    &\sim C_{H,0} + \frac{3 D_2 }{128\pi^2 G}\tilde{a}^3 H_I^2 \left(e^{3H_I t_I}-1 \right)\\
    & \sim \frac{3 D_2 }{128\pi^2 G} \tilde{a}^3 H_I^2 e^{3H_I t_I} = \frac{3 D_2 }{128\pi^2 G} a_I^3 H_I^2 
    \ ,
\end{split}
\end{equation}
where $C_{H,0}$ is an arbitrary initial violation of the constraint at the beginning of inflation, $a_I$ is the value of the scale factor at the end of inflation and in the last line we have taken the large $N_I$ limit. This shows that, if the inflationary phase lasts long enough, any initial violation of the constraint becomes negligible and we would be left at reheating with a positive average violation $C_H$. Combined with the previous result, that constraint violations act as pressureless dust, this shows that phantom cold dark matter can arise dynamically in a CQ theory of gravity.

A natural question is what variation around the mean one can expect for $C_H$ under the above assumptions. For this calculation, we take $C_{H,0} = 0$ since it is so easily swamped by the dynamically-generated constraint violation. First, consider:
\begin{equation}
\begin{split}
    \mathbb{E}[ C_H^2 ] = \bar{C}_I ^2 +\frac{27 D_2 H_I^5}{256\pi^3 G^2} \tilde{a}^6 \int_{0}^{t_I} \int_{0}^{t_I} e^{3H_I t'} e^{3H_I t''}  \mathbb{E}[ \bar{\zeta}(t') \bar{\zeta}(t'') ] \text{d}t'\text{d}t'' + \mathcal{O}(D_2^2) \ .
\end{split}
\end{equation}
Imposing \eqref{eq:noise_hub} one obtains (to linear order in the diffusion coefficient):
\begin{equation}
\begin{split}
    \sigma_{I}^2 &=  \mathbb{E}[ C_H^2(t_I)] -\bar{ C}_I ^2 \ , \\
    &=  \frac{9 D_2 }{512\pi^3 G^2} \tilde{a}^6 H_I^4 \left(e^{6H_I t_I}- 1\right) \\
    & \sim \frac{9 D_2 }{512\pi^3 G^2} \tilde{a}^6 H_I^4 e^{6H_I t_I} = \frac{9 D_2 }{512\pi^3 G^2} a_I^6 H_I^4 \ ,
\end{split}
\end{equation}
where again we have taken the large $e$-folds limit. Finally, one can evaluate the ratio between the standard deviation and the mean to be:
\begin{equation}
    \frac{\sigma_{I}}{\bar{C}_I} =  \frac{\sqrt{32\pi}}{\sqrt{ D_2}} \ ,
\end{equation}
independent of the number of $e$-folds during or after inflation.

The above results are derived assuming renormalization via averaging over a Hubble volume, such that the effective diffusion rate scales with $H^3$. For generality, we also report the result for a renormalisation procedure that leads to a polynomial scaling of arbitrary degree $n$ with the Hubble parameter $H$, to leading order in $D_2$. Although the deterministic drift for the Hubble parameter remains the Riccati equation with the same attractor, we are not aware of formal results for one-dimensional Riccati diffusion of that form. At any rate, for small diffusion, the nature of the drift guarantees that trajectories will be concentrated around the deterministic solution for appropriate initial conditions. From dimensional analysis, cosmic diffusion scaling with $H^n$ is related to the local diffusion coefficient by:
\begin{equation}
  \bar{D}_2 = \frac{3}{4\pi}D_2 H^{n} L^{n-3} \ , 
\end{equation}
up to numerical factors, where $L$ is some length scale needed for dimensional consistency. Under this general scaling the average constraint violation accumulated during inflation in the large $e$-fold limit is given by:
\begin{equation}
    \bar{C}_{I,n} \simeq  \frac{3 D_2 }{128\pi^2 G} a_I^3 H_I^{n-1} L^{n-3}  \ .
\end{equation}
Similarly, the variance amounts to:
\begin{equation}
    \sigma_{I,n}^2 \simeq  \frac{9 D_2 }{512\pi^3 G^2} a_I^6 H_I^{n+1} L^{n-3} \ ,
\end{equation}
meaning that the normalised variations have typical size:
\begin{equation}
    \frac{\sigma_{I,n}}{\bar{C}_{I,n}} =  \frac{\sqrt{32\pi}}{\sqrt{ D_2}}\left(H_I L\right)^{\frac{3-n}{2}} \ .
\end{equation}
For $n=3$ we indeed recover our previous result.

\subsection{Radiation domination and beyond}
In standard cosmological models, once inflation ends, the universe reheats into a radiation domination phase. During radiation domination, the violation of the Hamiltonian constraint continues to accumulate -- although its energy density is still subleading with respect to radiation. Since the comoving Hubble radius expands during radiation domination, one has the additional complication that patches which have different stochastic realisations during inflation now enter causal contact, necessitating an inhomogeneous calculation. 

However, the rate of change for $C_H$ depends cubically on $H$ when averaging over a horizon patch. $H$ drops linearly with time during radiation domination, so phantom cold dark matter is generated significantly only in the first few $e$-folds of radiation domination, where one can still work in the separate Universe approximation. By the time significant inhomogeneities enter the horizon, the noise has effectively decoupled from the evolution of the scale factor. From there on, we can treat the Universe as satisfying the standard Friedmann equations, with the density (and density fluctuations) of phantom cold dark matter already determined. As the universe later evolves out of radiation domination and into matter domination, this picture continues to hold since $H$ continues to drop.

To evaluate the phantom cold dark matter produced during the early stages of radiation domination, we can repeat the same calculation as before, now assuming that the zeroth order evolution in the scale factor is given by:
\begin{equation}
    a(t) = a_I \sqrt{2 H_I t + 1} \ ,
\end{equation}
where we have matched both the scale factor and the Hubble parameter at the end of inflation with the respective quantities at the beginning of radiation domination. As before, this can be made precise by considering the evolution of the Hubble parameter
\begin{equation}
    \dot{H} = -2H^2 - \sqrt{\frac{3}{16\pi} H^3 D_2}\bar{\zeta} \ ,
\end{equation}
elucidating that $1/H$ is a squared Bessel process. It is a matter of simple algebra to see that, while at first the fractional deviation in the process increases as $t^{1/2}$, at large times it saturates to
\begin{equation}
    \frac{\sqrt{\textrm{Var($H$)}}}{\mathbb{E}[H]} \to \sqrt{\frac{1}{2+ \frac{64\pi}{3D_2}}} \approx \sqrt{\frac{3 D_2}{64\pi}} \ll 1\ ,
\end{equation}
meaning the trajectories for the Hubble parameter are concentrated around the deterministic trajectory for radiation domination as well.

Note, we are re-shifting time such that radiation domination runs from $t=0$ to $t=t_R$. We indicate with $\bar{ C}_{R}$ the average violation of the Hamiltonian constraint accumulated during radiation domination:
\begin{equation}
\begin{split}
    \bar{C}_R&= \frac{9 D_2}{128 \pi^2 G}a_I^3 \int_{0}^{t_R} \left( H  \sqrt{2 H_I t' + 1} \right)^3  \text{d}t'\\
    & = \frac{9 D_2}{128 \pi^2 G}a_I^3 H_I^3 \int_{0}^{t_R} \left(2 H_I t' + 1\right)^{-3/2}  \text{d}t'\\
    & = \frac{9 D_2}{128 \pi^2 G}a_I^3 H_I^2 \left(1- \frac{1}{\sqrt{2 H_I t_R +1}} \right) \\
    & \sim  \frac{9 D_2}{128 \pi^2 G}a_I^3 H_I^2 = 3 \bar{ C}_I \ .
\end{split}
\end{equation}
In the last line, we have dropped the $t_R$-dependent term since we have taken the large $N_R$ limit for radiation domination as well. As expected, most of the phantom matter density is accumulated in the first few $e$-folds of radiation domination. That leads to a factor of 3 enhancement with respect to what was generated during inflation, i.e.:
\begin{equation}
    \frac{\bar{C}_R}{\bar{C}_I } = 3 \ .
\end{equation}
We can similarly calculate the growth in the size of the fluctuations during radiation domination. The variance of the violation is given by:
\begin{equation}
\begin{split}
    \sigma^2_R &= \frac{27}{256\pi^2 G^2}a_I^6 D_2 H_I^2\int_0^{t_R}  \int_0^{t_R} H^3 \mathbb{E}[ \bar{\zeta}(t') \bar{\zeta}(t'') ] \times \\
    & \hspace{100pt} \times  \sqrt{2H_I t'+1}\sqrt{2H_I t''+1} \text{d}t' \text{d}t'' \\
    & = \frac{27}{256\pi^2 G^2} a_I^6 H_I^5 D_2  \frac{t_R}{2H_I t_R + 1} \\
    & \sim \frac{27}{512\pi^2 G^2} a_I^6 H_I^4 D_2 = 3\sigma^2_I \ .
\end{split}
\end{equation}
The total final variance is additive, since the noise is uncorrelated in time. Therefore:
\begin{equation}
    \sigma^2 = \sigma^2_I + \sigma^2_R
\end{equation}
Hence, at the end of radiation domination, the density contrast in $\bar{C}_H $ is given by:
\begin{equation}
\label{eq:after_rad}
    \frac{\sigma}{\bar{C}_H } = \frac{\sqrt{\sigma_I^2 + \sigma^2_R}}{\bar{C}_I + \bar{C}_R} = \frac{1}{2} \frac{\sigma_I}{\bar{C}_I} = \frac{\sqrt{8\pi}}{\sqrt{ D_2}} \ .
\end{equation}

Since we have been working in the limit of $\sqrt{D_2} \ll 1$, $\sigma_I$ would be much greater than the average violation, leading to negative densities in some regions. At first sight this is a major problem, since measurement of the CMB temperature fluctuations show $\delta T/T \ll 1$, which, in standard cosmology, is a quantity related to the density contrast $\delta \rho/\rho$ of matter. However, \eqref{eq:after_rad} is not the quantity that we expect to observe in the CMB itself, because it is strongly dominated by fluctuations on microscopic scales (in the order of the comoving horizon scale at the end of inflation). Indeed, there are even more extreme fluctuations that have been averaged over to write down an effective homogeneous theory. The evolution of these microscopic scales is beyond the scope of this article, but we continue to work on the assumption that cosmological effects will see an effective density averaged over relevant scales. 
 If this is the case, the variance in the phantom dark matter density will be drastically dampened by a factor given by the averaging volume, yielding a positive energy density with very small perturbations around the mean.  
 
 On top of these perturbations, one must consider the effect of quantum fluctuations in the scalar field that drives inflation, to obtain a full picture of the inhomogeneities during radiation domination. This calculation will require use of the local theory rather than the averaged one, since one expects {\it concentration of measure} effects which may depend on the density of states in the local theory. We therefore leave the important question of observable density fluctuations to future work.

Returning to the mean density, we should consider the effect of our renormalisation choice on the result. For general polynomial scaling with $H$, the average constraint violation accumulated during radiation domination is:
\begin{equation}
    \bar{C}_R =  \frac{9 D_2}{128 \pi^2 G}H_I^{n-1}L^{n-3} \frac{(2 H_I t_r +1)^{\frac{5-2n}{2}}-1}{5-2n} a_I^3 \ .
\end{equation}
Whilst for inflation different $n$ trivially translated into different scaling of the final result with $H_I$, in radiation domination the situation is more complex, due to the dynamical nature of the Hubble parameter. Indeed, the accumulation of constraint violation continues long into radiation domination unless $n\geq 3$. This would pose severe issues for interpreting the stochastic effects as phantom cold dark matter, since the gravitating density would vary while baryon acoustic waves propagated through the early universe, likely violating CMB constraints~\cite{akrami2020planck}.

\subsection{Estimating the amount of phantom dark matter}

The energy scale of inflation, the $e$-folds of inflationary expansion and the averaged diffusion coefficient of the theory determine the amount of phantom dark matter generated. To check whether observational bounds on the cosmological density parameters can rule out such a mechanism for dark matter generation, we compute the density parameter of phantom dark matter today given that we generate phantom dark matter with average energy density $\bar{C}_H$ in the early stages of radiation domination. As usual, we define the density parameter of an energy species $i$ as:
\begin{equation}
    \Omega_i = \frac{\rho_i}{\rho_c} \ ,
\end{equation}
where $\rho_c = 3 H_0^2 /8\pi G$ is the critical energy density. Then:
\begin{equation}
    \Omega_c(a) = \frac{8\pi G \bar{C}_H a^{-3}}{3 H_0^2} =  \frac{4 D_2 }{\pi}\frac{H_I^2}{H_0^2}\frac{a_I^{3}}{a^{3}} \ ,
\end{equation}
which, assuming $N$, $P$ and $M$ $e$-folds of inflation, radiation domination and matter domination respectively, reduces to:
\begin{equation}
\label{eq:omegac_est}
     \Omega_c(a) =  \frac{4 D_2 }{\pi} \frac{H_I^2}{H_0^2} e^{-3(M+P)} \ .
\end{equation}
The $e$-folds of inflation naturally drop from the expression since the generation of dark matter and its dilution due to the expansion of the Universe have opposite scaling with the scale factor $a$. However, $H_I/H_0$ is completely determined once $M$ and $P$ are known. Indeed, at the end of radiation domination we have:
\begin{equation}
    H_R = H_I e^{-2P} \ ,
\end{equation}
whilst from the beginning of matter domination to today the Hubble parameter evolves to:
\begin{equation}
    H_0 = H_R e^{-3M/2} = H_I e^{-\frac{4P + 3M}{2}} \ .
\end{equation}
Plugging this back into \eqref{eq:omegac_est} we have that the $e$-folds of matter domination also drop from the expression, leaving:
\begin{equation}
     \Omega_c(a) = \frac{4 D_2 }{\pi}  e^{P} \ .
\end{equation} 

The phantom cold dark matter density thus depends only on the dimensionless diffusion constant $D_2$ and $P$, the number of $e$-folds of radiation domination, meaning that it is in principle determined by existing cosmological constraints and laboratory limits on diffusion.
$P$ is constrained by the ratio of the temperature at matter-radiation equality ($z_{eq}=3400$) and at the temperature after reheating, which in turn is related to the inflationary energy (often taken to be GUT scale). Minimal models of inflation usually constrain $P\approx 55$, following from bounds on the inflation energy scale given by CMB data~\cite{akrami2020planck}. 

Table-top experiments bound the value of the local diffusion coefficient $D_2$. Relating the diffusion at the energy scales of terrestrial experiments, to the higher energy scale of inflation requires, other than the averaging procedure already described and the renormalisation of $D_2$ that accompanies it, a careful consideration of the RG flow of the stochastic theory at different energy scales~\cite{grudka2024renormalisation}. Since the theory can be related to quadratic gravity which is asymptotically free, the diffusion constant $D_2$ is expected to run~\cite{julve1978quantum,fradkin1981renormalizable, benedetti2009asymptotic,buccio2024physical,grudka2024renormalisation}. The stochastic fluctuations may also have an effective mass~\cite{grudka2024renormalisation}, which can also suppress fluctuations at lower energy scales. Finally, preliminary results indicates that the effective diffusion coefficients depend on two free parameter, which can take values that enhance fluctuations of the Hamiltonian constraint, while keeping the random forces in the Newtonian potential small.  These topics are still active area of research, and most of the questions are still open, meaning that we cannot currently perform that mapping reliably. The best we can do at the moment is to use current bounds on the diffusion coefficient from table-top experiments and assume that they trivially apply in \eqref{eq:omegac_est}, in order to showcase how the combination of cosmological observations and table-top experiments can provide a powerful stress-test for classical-quantum theories of gravity.  

With some assumptions, data from LISA Pathfinder constrains the dimensionless diffusion coefficient of scalar fluctuations to be $D_2\lessapprox 10^{-54}$ in the case the fluctuations have no effective mass ~\cite{grudka2024renormalisation}.
Another bound could, in principle, be extracted from the production of stochastic gravitational waves and LIGO data -- although doing this relies on additional modelling, needed to relate the noise in the transverse-traceless modes to the fundamental constant $D_2$ for scalar modes, which requires fixing a free parameter~\cite{grudka2024renormalisation,oppenheim2025diffusion}. Assuming $P=55$ $e$-folds of radiation domination and saturating such bound coming from the Pathfinder mission, we obtain $\Omega_c \lessapprox 10^{-31}$ as the density parameter of phantom dark matter today. In order to obtain an $\Omega_c$ of order unity to explain all of the observed dark matter density, one would require either $P\approx 124$ $e$-folds of radiation domination or, keeping $P=55$, a diffusion coefficient of $D_2\approx 10^{-24}$ -- several orders of magnitudes above the experimental bound.

This back-of-the-envelope calculation has to be taken lightly since, as discussed above, flowing the local theory to the IR cosmology consistently is a non-trivial step, currently a work in progress. Still, if, like in the case we have discussed, $\bar{C}_H$ turns out to be very small, then one cannot rule out hybrid gravity since CDM can always be included as a bona fide matter component in the Universe. The upshot would be that the cosmological departure from the constraint would be very hard to detect, and the cosmology is well approximated by deterministic GR models even if spacetime is indeed fundamentally classical. If instead precise measurement of $H_I$, $P$ and $D_2$ lead to a larger-than-observed amount of phantom CDM, one would rule out CQ theories that do not preserve the deterministic constraints of GR. Indeed, even though the GR constraint violation is a natural consequence of the model we have considered, one can expect that non-minimal models exist in which the constraints are satisfied exactly at the level of trajectories. We describe such an option in Section \ref{sec:discussion} and, more in detail, in Appendix \ref{app:constraint}. In general, however, the expectation is that such dynamics require a modification of the equations of motion even at the level of the deterministic drift, which makes them less desirable. In the middle of these two extremes is the possibility that phantom CDM is produced with the correct density to account for cosmological CDM; however, the calculations above show that this would imply a fine-tuned relationship between $P$ and $D_2$. 

Although popular, not all cosmological models include an inflationary epoch. Indeed, bouncing universes have been proposed as alternative solution to the horizon problem, sidestepping the issue of the primordial cosmological singularity~\cite{friedmann1922curvature,tolman1931periodic}, with mixed success~\cite{battefeld2014critical,brandenberger2016bouncing}. Bounces are common in loop quantum cosmology~\cite{ashtekar2006quantum}, but they can arise also classically in the presence of either positive spatial curvature or exotic matter~\cite{battefeld2014critical}. Assuming, as we have done throughout the article, that the energy density due to the stochastic fluctuation is always subleading, our analysis goes through intact in a bouncing Universe scenario too (although one needs to be careful about the definition of the causal patch of averaging, as the one we have adopted in this article breaks down at the bounce, where $H=0$). Indeed, energy is produced in any FLRW stochastic cosmology, irrespective of the specific fluid sourcing it -- the size of the constraint violation generated depends on the specific bouncing model. The constraint violation produced prior to the bounce will then add to the one later generated in the early stages of radiation domination, for which the computation we have just described remains valid. 

\section{Discussion} \label{sec:discussion}
\subsection{Summary}
Theories of gravity that aim to reconcile quantum mechanics and general relativity without quantising spacetime must necessarily be stochastic in the gravitational degrees of freedom. We now briefly review our results, before highlighting the open questions, discussing future research directions and similarities with other stochastic models of gravity.

We have shown that a natural consequence of the stochasticity is the violation of the deterministic constraints of GR. We argued why in the low-noise limit, such a constraint violation acts as an effective  pressureless dust both in the homogeneous and inhomogeneous treatment. Then, we presented a mechanism by which, during inflation, the integrated effect of the stochasticity leads to a constraint violation that is, on average, positive. This provides one of the missing ingredients needed to show that such an effect could imitate the cosmological fingerprint of cold dark matter. 

We discussed how, for a natural choice of renormalisation scheme in which the homogeneous global noise is related to the local one by averaging across the Hubble horizon, the phantom matter generation stops a few $e$-folds after inflation, since the effective diffusion coefficient drops as $H^3\sim t^{-3}$. This is a necessary condition for this mechanism to produce phantom cold dark matter that is consistent with CMB constraints. Shortly after the beginning of radiation domination, the cosmological noise becomes negligible and the evolution is adequately captured by the standard Friedmann's equations with phantom CDM on cosmological scales. The average density of the phantom matter follows a simple relation that depends on the horizon-scale diffusion coefficient, the energy scale of inflation and the number of $e$-folds of expansion of the Universe. Improving existing constraints on $H_I$, the number of $e$-folds of inflation and the inferred CDM density will in turn put tight constraints on $D_2$. If details of the renormalization procedure can be worked out, one could test for consistency between the horizon-scale $D_2$ implied by cosmological observations and constraints coming from table-top experiments of the local $D_2$~\cite{oppenheim2023gravitationally}.

\subsection{Comparison with other models}
The motivation behind our approach is that stochasticity in the classical gravitational degrees of freedom is a necessary consequence of a theory that describes classical and quantum systems with non-trivial interaction in a consistent manner. In particular, it follows once the physical requirements of complete positivity, trace preservation and linearity are imposed at the level of the master equation, which is the evolution equation of the hybrid state~\cite{oppenheim2021postquantum,oppenheim2022classes}. While this specific motivation has emerged over the past few years, having some randomness in the evolution equations for the gravitational field is not a new proposal, with stochastic evolution equations put forward mainly as effective theories. Here we give a brief overview of some of these approaches and how they differ from our analysis.

Cosmetically, the closest approach to our own comes from the formalism of “stochastic gravity"~\cite{hu2008stochastic}. Its objective is to include quantum fluctuations when calculating backreaction effects in semiclassical gravity. In stochastic gravity, the stochastic tensor $\xi_{\mu \nu}$ represents the quantum fluctuations of the field that sources the metric and is therefore not a fundamental white noise process. Indeed, the moments of the random field match those of the stress energy tensor of the QFT. Stochastic gravity is governed by the Einstein-Langevin equations 
\begin{equation}
    G_{\mu\nu} = 8\pi G \left(\langle  T_{\mu\nu} \rangle + \xi_{\mu\nu}\right),\label{eq:semi}
\end{equation}
and aims to include corrections to the semiclassical Einstein's equations, with which it partially shares the regime of validity (i.e. when the fluctuation in the stress-energy tensor of the quantum system are small with respect to the mean). This condition is commonly probed by the Kuo-Ford criterion~\cite{kuo1993semiclassical}. Indeed, even though stochastic gravity can improve on the semiclassical Einstein's equation by including both the quantum fluctuations of the matter fields and the induced one of the gravitational field~\cite{ahmed2024semiclassical}, the expectation value in Eq. \eqref{eq:semi} leads to a breakdown in causality unless the theory is modified in some way~\cite{gisin1990weinbergs,polchinski1991weinbergs} due to the non-linearity in the density matrix, much like in the standard semiclassical Einstein's equations. As such, stochastic gravity can only be treated as an effective theory. This is also true for the other proposals of stochastic theories of gravity that modify Einstein's equations by adding a random gravitational constant $G$ instead~\cite{moffat1997stochastic,decesare2016effective}. Stochastic gravity and these proposals share the feature that their starting point is the full set of Einstein's equations, with the constraints and dynamical part fluctuating in the same way as is clear from \eqref{eq:semi}. 

A virtue of CQ gravity, both in its covariant path integral formulation and the $3+1$ master equation approach, is that the dynamics is linear in the hybrid classical-quantum state. This allows for encoding correlations between gravitational and matter degrees of freedom, meaning that the formalism can be applied beyond effective regimes, much in the same way that studying quantum mechanics at the level of density matrices allows for the support of both classical and quantum correlations. Indeed, the CQ framework can be seen as a generalisation of the continuous measurement-and-feedback formalism. The gravitational field acts as a measurement device and, conditional on the “outcome", the quantum state consequently decoheres, correlating the classical and quantum fields with each other. 

Causal set theory and unimodular gravity also motivate some models of effective spacetime diffusion. In the everpresent Lambda proposal~\cite{ahmed2004everpresent}, the stochasticity in the cosmological constant comes from the causal set interpretation of $\Lambda$ being the conjugate variable to the local spacetime volume, a stochastic variable. In unimodular gravity (often considered the continuum limit of causal set theory) the motivation of having a stochastic evolution of spacetime comes from violation of the conservation of the matter stress-energy tensor instead, which is allowed by the theory~\cite{perez2021resolving,landau2022cosmological}. 

\subsection{Open questions}
The scheme in this work allows for the violation of the constraints of general relativity. 
As mentioned in the introduction, constraints are a necessary condition for covariance in the deterministic theory. In the Hamiltonian ADM formulation, they are required for the series of three-dimensional spatial geometries to be embeddable into a (3+1)-dimensional spacetime. However, in a stochastic theory the role of constraints is more subtle and their violation should not be taken to indicate loss of covariance, but rather that the constraint was formulated ignoring the existence of the stochastic field. Indeed, the two approaches typically used to derive constraints~\cite{dirac1967lectures,hojman1976geometrodynamics} in the deterministic theory fail here. The Dirac procedure is not applicable as the Hamiltonian is not the generator of the dynamics, a point already made in~\cite{oppenheim2021constraints}. The expectation is that the concept of the  constraint needs modifying for random systems, similarly to what happens for Noether's theorem~\cite{misawa1999conserved, albeverio1995remark,oppenheim2009fundamental}. Moreover, deriving the constraints by demanding that the hypersurface deformation algebra closes~\cite{hojman1976geometrodynamics}, yields no constraints in minisuperspace since the algebra is trivial. Of course, the situation is more complex when considering the full local theory, as discussed in~\cite{oppenheim2021constraints}. Diffeomorphism invariance then hinges on the closure of the constraint algebra, here generated by stochastic operators. Alternatively, it has been shown that stochastic theories of gravity, as well as classical-quantum theories of gravity, can indeed be covariant via the path-integral formulation~\cite{oppenheim2023covariant}, for example, in the context of Nordstrom gravity~\cite{oppenheim2024diffeomorphism}. However it is harder to confirm whether this formulation is completely positive and trace-preserving (CPTP). Conversely, the continuous master-equation of~\cite{oppenheim2021postquantum} is CPTP, but it is not yet known if it is covariant or equivalent to the path integral approach. It has been shown that models with discrete jumps in the classical degrees of freedom cannot be covariant~\cite{oppenheim2021constraints}, since there the hybrid constraints cannot be satisfied.

Our construction is indeed valid only as a description within a preferred frame, the cosmological one, since spacetime looks homogeneous and isotropic only within a specific set of coordinates. These, adapted to the symmetries of the problem, are effectively provided by the perfect fluid sourcing the geometry. It is only in these coordinates that we can identify a consistent low-noise regime on late-time spatial hypersurfaces. On a different, arbitrary, foliation of spacetime, the homogeneous description breaks down and a more refined analysis is needed.

For completeness, we note that it is possible to construct stochastic dynamics that satisfy the GR constraint exactly, although we expect such a theory to violate diffeomorphism invariance. To construct it, it is necessary to incorporate noise into the evolution law for $\rho$ as well as $\pi_a$, as we show in Appendix \ref{app:constraint}. This requires a nontrivial coupling of the noise field with matter, which goes beyond the scope of this study. We also show that the GR constraint cannot be enforced by making stochastic the equation of motion of the scale factor instead.

This work sets the basis for the study of the cosmological consequences of theories where gravity is classical and, therefore, fundamentally stochastic. The main objective of this line of work is to try to establish predictions that may help test such theories in the near future, possibly proving observationally the necessity of quantising the gravitational theory. CQ theories of gravity predict that the amount of diffusion in the gravitational system is lower-bounded by the typical decoherence rate of superposition in the quantum matter degrees of freedom. As explored in detail in~\cite{oppenheim2023gravitationally}, table-top experiments can lower-bound the diffusion in the metric and squeeze the theory from both sides.

Here, we find that cosmological measurements can also provide easily accessible data that can also constrain the amount of diffusion allowed, on a very different scale. Indeed, as showcased in this article, when integrated over long times, the stochastic force can cause the gravitational system to diverge significantly from its deterministic trajectory. In particular, we have seen that if CDM can be understood by the mechanism presented in this article, its abundance would be entirely fixed by the parameters of the early-Universe model.

It is possible in principle to relate table-top precision tests of gravity to cosmological data. However, in order to do that, we need a better understanding of a number of theoretical issues. One important ingredient towards this analysis will be to understand how the global diffusion in the evolution of the scale factor in an FLRW model and at high energies, is related to the local one measured in the lab at low energies.
This requires a deeper understanding of the renormalisation group flow of the stochastic theory, a task initiated in~\cite{grudka2024renormalisation}. In particular, we want to understand how to flow the diffusion coefficients to high energy, such that it is possible to relate the $D_2$ appearing in \eqref{eq:omegac_est} to the one that is measured in table-top experiments. In addition to the running of the coupling constant $D_2$, we also need to understand if the fluctuations have mass, which would suppress fluctuations at low energy.
We would also like to relate this dynamical mechanism which generates phantom cold dark matter, to the static case considered in~\cite{oppenheim2024anomalous}. There, it was found that we should expect anomalous contributions to the deterministic weak field solution, which could act like dark matter. If the phantom CDM generated is cosmologically significant, the system would be driven locally to metrics that reproduce galactic Dark Matter phenomenology. Conditioned on this fact, the path integral representation of CQ gravity would favour (on galactic scales) a cosmological constant term that agrees with current estimates of $\Lambda$~\cite{oppenheim2024anomalous}.

Another key step towards using cosmological data to constrain the value of the diffusion coefficient in stochastic theories of gravity is studying the inhomogeneous evolution. This is key to reconstruct clear predictions for the observed Universe and especially the CMB; we know the separate-universe approximation is not really self-consistent, since the shrinking horizon during inflation partitions the universe into multiple causally-disconnected regions. The resulting inhomogeneous modes will re-enter the horizon following reheating, and their subsequent evolution can only be handled via an inhomogeneous calculation. Furthermore, whilst the homogeneous noise at late times goes to zero, locally the dynamics is still stochastic, meaning that late-time evolution might still differ from standard deterministic calculations. Altogether, moving away from the homogeneous model is fundamental in order to extract the power spectrum of the perturbations imprinted on the CMB that the stochastic theory predicts. This will ultimately provide a powerful stress-test for CQ theories of gravity, since there is a plethora of strong observational constraints that the stochastic theory will have to reproduce~\cite{aghanim2020planck}.

\vspace{3mm} 

\noindent\section*{Acknowledgements}

The authors would like to thank Christian Boehmer, Thomas Colas, Alex Jenkins and Benjamin Joachimi for helpful comments. EP is supported by the Cosmoparticle Initiative at University College London.
EP would like to thank Maite Arcos, Isaac Layton, Andrea Russo, Andrew Svesko and Zachary Weller-Davies for invaluable support and useful discussions. AP is supported by the European Union's Horizon 2020 research and innovation programme under grant agreement No. 818085 GMGalaxies. JO was supported by the Simons Foundation, It from Qubit Network.
\newpage

\appendix

\section{Noise and time reparametrisation}
\label{app:noise}
To study the evolution of the constraint under the modified equations of motion, we need to specify the white noise $\xi$ in terms of Wiener processes. A Wiener process (or, equivalently, Brownian motion) $W_t$ is an almost surely continuous stochastic process with Gaussian increments distributed as:
\begin{equation}
    \Delta W = W_t - W_{t'} \sim \mathcal{N}(0,t-t') \ .
\end{equation}
Schematically, the infinitesimal difference of a Wiener process $\text{d}W_t$ is of order $\mathcal{O}(\sqrt{\text{d}t})$, a statement made precise via It\^{o}'s lemma (see Appendix \ref{app:ito} and \ref{app:constraint} for more detail). In the physics community, it is standard to define the white noise field as the distributional derivative of Brownian motion~\cite{Oksendal:2023aa}, i.e:
\begin{equation}
    \xi \sim \frac{\text{d}W_t}{\text{d}t} \ .
\end{equation} 
However, $\xi$ so defined fails to be a scalar under time reparametrisation (it is a density of weight 1/2), which can be seen from dimensional analysis. The generalisation needed is readily  given by:
\begin{equation}
    \xi \equiv \frac{\text{d}W_t}{\text{d}t} \to \xi \equiv \frac{1}{N}\frac{\text{d}W_{\int N\text{d}t}}{\text{d}t} = \frac{1}{\sqrt{N}}\frac{\text{d}W_t}{\text{d}t}  \label{eq:noise}
\end{equation}
which has the required transformation properties and inherits from the Wiener process the statistics postulated in \eqref{eq:noise_gen} as we now show.

After a monotonic redefinition of time $u=g(t)$, we have that $\text{d}W_t$ can be expressed in terms of a Wiener process with respect to the new time $u$ as:
\begin{equation}
    \mathrm{d}W_{u=g(t)}= \sqrt{\dot{g}} \  \mathrm{d}W_{t} \ .
\end{equation}
This can be understood both from the scaling property of a Wiener process and from It\^{o}'s lemma ($\text{d}W_t \sim \sqrt{\text{d}t}$). Therefore, the stochastic integral:
\begin{equation}
    \int \Sigma[N] \text{d}W_t \ 
\end{equation}
is left invariant under such a transformation if $\Sigma$ to transforms under time reparametrisation as:
\begin{equation}
    \Sigma' = \frac{1}{\sqrt{\dot{g}}} \Sigma \ .
\end{equation}
Hence, we need $\Sigma$ to be a scalar density in time of weight $w=1/2$. In this form, it is immediate to see that the choice $\Sigma[N]= \sqrt{N}$ gives the desired transformation property. The invariant stochastic integral takes the form:
\begin{equation}
    \int \sqrt{N_t} \text{d}W_t = \int \xi N \text{d}t \ ,
\end{equation}
where the derivative of the Wiener process has to be understood in a distributional sense.

We can now trivially extract the moments that the stochastic field inherits from the following properties of the Wiener process:
\begin{equation}
    \langle W_t \rangle=0 \ , \qquad \langle W_t W_s \rangle = \textrm{min}(s,t) \ ,
\end{equation}
which, as we now show, lead to:
\begin{equation}
    \mathbb{E}[ \xi(t)] =0 \ , \qquad  \mathbb{E}[ \xi(t)\xi(t') ] = \frac{1}{N(t)}\delta(t,t'). \
\end{equation}
First, consider the definition of $\xi$ as a distributional derivative of the Wiener process acting on some test function $\phi$:
\begin{equation}
\begin{split}
     \xi[\phi] &= \int_0^{\infty} \xi(t) \phi(t) N(t) \mathrm{d}t \\
    & = -\int_0^{\infty} W(t) \frac{\mathrm{d}}{\mathrm{d}t}\left( \sqrt{N(t)} \phi(t) \right) \mathrm{d}t \ .
\end{split}
\end{equation}
Then:
\begin{equation}
\begin{split}
    \mathbb{E}[ \xi[\phi] ] &= - \int_0^{\infty}\mathbb{E}[ W(t) ] \frac{\mathrm{d}}{\mathrm{d}t}\left( \sqrt{N(t)} \phi(t) \right) \mathrm{d}t = 0 \\
    &= \int_0^{\infty} \mathbb{E}[  \xi(t)] \phi(t) N(t) \mathrm{d}t \ , 
\end{split}
\end{equation}
yielding $\mathbb{E}[\xi] = 0$.
For the variance, start from:
\begin{equation}
\begin{split}
    \mathbb{E}[ \xi[\phi]^2] & = \int_0^{\infty} \int_0^{\infty} \mathbb{E}[ W(t) W(t')] \frac{\mathrm{d}}{\mathrm{d}t}\left( \sqrt{N(t)} \phi(t) \right) \mathrm{d}t \frac{\mathrm{d}}{\mathrm{d}t'}\left( \sqrt{N(t')} \phi(t') \right) \mathrm{d}t'  \\
    & = \int_0^{\infty} \int_0^{\infty} \mathrm{min}(t,t') \frac{\mathrm{d}}{\mathrm{d}t}\left( \sqrt{N(t)} \phi(t) \right) \mathrm{d}t \frac{\mathrm{d}}{\mathrm{d}t'}\left( \sqrt{N(t')} \phi(t') \right) \mathrm{d}t' \\
    & = \int_0^{\infty} \frac{\mathrm{d}}{\mathrm{d}t}\left( \sqrt{N(t)} \phi(t) \right) \left[ \int_0^{t}\sqrt{N(t')} \phi(t') \mathrm{d}t' \right] \mathrm{d}t\\
    & = \int_0^{\infty} \int_0^{\infty} \delta(t,t') \sqrt{N(t)} \phi(t) \sqrt{N(t')} \phi(t') \ \mathrm{d}t' \mathrm{d}t\\.
\end{split}
\end{equation}
However, we also have:
\begin{equation}
    \mathbb{E}[ \xi[\phi]^2 ] = \int_0^{\infty} \int_0^{\infty}  \mathbb{E}[ \xi(t) \xi(t') ] \ \phi(t) \phi(t') N(t) N(t') \ \mathrm{d}t \mathrm{d}t',
\end{equation}
and comparing the two we obtain the claimed variance for the noise process, up to an arbitrary scale $D_2$.

\section{It\^o vs Stratonovich} \label{app:ito}
Whilst this is by no means intended to be a comprehensive review on stochastic calculus (for one without too much mathematical baggage, see~\cite{risken1985fokker}), here we discuss briefly the difference between It\^{o} and Stratonovich integration to motivate our choice of the former over the latter in interpreting \eqref{eq:eom}.

When it comes to stochastic differential equations, their interpretation is not  unique. The reason is that, unlike Riemann integrals, stochastic integrals give different results for different discretisations. That is, Riemann integrals:
\begin{equation}
    \int_0^t f \text{d}t = \lim_{n\to \infty} \sum_{[t_{i+1},t_i] \in \pi_n } f(\alpha Z_{t_i} + (1-\alpha) Z_{t_{i+1}}) (t_{i+1}-t_i) \ ,
\end{equation}
where $\pi_n$ is a sequence of partition of $[0,t]$ with mesh going to zero, are equivalent for every $\alpha \in [0,1]$ as $\delta t \to 0$. In contrast, a stochastic integral defined similarly as:
\begin{equation}
    \int_0^t f \text{d}W_t = \lim_{n\to \infty} \sum_{[t_{i+1},t_i] \in \pi_n } f(\alpha Z_{t_i} + (1-\alpha) Z_{t_{i+1}}) (W_{i+1}-W_i) \ ,
\end{equation}
where $W_t$ is a Wiener process, gives different results for different choices of $\alpha$. Consider the following stochastic differential equation:
\begin{equation}
\label{eq:sde_appA}
    \text{d}Z = \mu(Z) \text{d}t + \sigma(Z) \text{d}W_t \ ,
\end{equation}
with $\sigma$ and $\mu$ being related to the variance and the average drift of the stochastic process respectively. Different definitions of stochastic integration lead to different interpretation of the differential equation.
Choosing $\alpha=0$, the update of the state $Z_t$ from $t$ to $t+\Delta t$ is given by:
\begin{equation}
    Z_{t+\Delta t} =Z_t + \mu(Z_t) \Delta t +\sigma(Z_t) \Delta W_t \ .
\end{equation}
This is known as It\^o integration. It is non-anticipative (i.e. the increment is evaluated with information of the functions $\mu$ and $\sigma$ at the current time-step only) and it is easier to handle numerically, but it does not obey the standard chain rule. Instead, the total derivative of a function of a stochastic variable $Z_t$ (evolving under \eqref{eq:sde_appA}) obeys It\^{o}'s lemma instead:
\begin{equation}
    \text{d}f(Z_t,t) = \left(\frac{\partial f}{\partial t} + \frac{\partial f}{\partial z} \mu \right) \text{d}t + \frac{\partial f}{\partial z} \sigma \text{d}W_t + \frac{1}{2} \frac{\partial^2 f}{\partial z^2} \sigma^2 \text{d}t \ .
\end{equation}
The usual interpretation of the extra $\text{d}t$ term appearing in the formula for the total derivative is that the Wiener process is of fractional order in time. In particular, $\text{d}W_t^2 \sim \text{d}t$.

On the other hand, $\alpha=1/2$ leads to:
\begin{equation}
    \Delta Z_t = f\left(\frac{Z_t+Z_{t+\Delta t}}{2}\right) \Delta t +\sigma\left(\frac{Z_t+Z_{t+\Delta t}}{2}\right) \Delta W_t \ .
\end{equation}
This has an obvious generalisation to higher dimensional degrees of freedom. This choice is known as Stratonovich interpretation and does respect the usual chain rule, but the background noise is anticipative. Stratonovich definition appears naturally when deriving stochastic differential equations through coarse-graining. Therefore, the latter is commonly used in stochastic physical systems, where the noise is introduced as the effective description of some environment~\cite{kampen1981ito}. 

Whilst Stratonovich and It\^{o} calculus counterintuitively yield different result for the same differential equation, they only differ by a drift term. Indeed, one can move between an It\^o and a Stratonovich intepretation by correcting the drift term $\mu$ in the equations of motion in such a way to cancel the spurious extra $\mathcal{O}(\text{d}t)$ contribution in It\^{o}'s lemma. Hence, if a system is described by a Stratonovich equation, it can be translated into an It\^o system and, therefore, easily simulated on a computer.

\section{Inhomogeneous evolution} \label{app:inhomogeneous}
Consider a homogeneous cosmology with small perturbations. Expanding the metric to linear order as $g_{\mu\nu} = g^{(0)}_{\mu\nu}+ \epsilon h_{\mu\nu}$, where $\epsilon$ is a small parameter and $g^{(0)}_{\mu\nu}$ satisfies the background Friedmann's equations sourced by a homogeneous perfect fluid $T^{(0)}$ with density $\rho$ and equation of state parameter $w$. For simplicity, we consider a single background fluid, but the argument is insensitive to this assumption. The homogeneous background evolution can also be off the constraint surface by $C_H$. The metric perturbations solve the standard linearised equations for cosmological perturbations. For simplicity, consider the cosmological perturbations in Newtonian gauge:
\begin{equation}
\begin{split}
    \text{d}s^2 = a^2(\tau) \Big[& -\left(1+2 \epsilon \phi(k,\tau) e^{ik^i x_j} \right) \text{d}\tau^2 \\
    & + \left(1-2 \epsilon \psi(k,\tau) e^{ik^i x_j} \right) \delta_{ij} \text{d}x^i \text{d}x^j \Big] \ ,
\end{split}
\end{equation}
where we have expanded the metric functions in Fourier modes, since their equations of motion will be linear, and have assumed $k=0$ for the background solution. These metric perturbations are sourced by small inhomogeneous perturbations in the matter density. For a comoving perfect fluid, these are given by:
\begin{equation}
\begin{split}
    \delta T^{0}_0 &= \epsilon \delta \rho \ , \\
    \delta T^{0}_i &= \frac{\epsilon}{a}(1+w) \rho  \delta U_i \ , \\
    \delta T^{i}_j &= - \epsilon w \delta \rho \ \delta^i_j \ , 
\end{split}
\end{equation}
where $\delta \rho$ is the perturbation in energy density in the matter field, whilst $\delta U^i$ is the relative velocity of the fluid perturbations with respect to the comoving frame. Similarly, the  inhomogeneous perturbations to the global constraint violation read: 
\begin{equation}
\begin{split}
    \delta C^{0}_0 &=\epsilon \delta C \ , \\ 
    \delta T^{0}_i &= \frac{\epsilon}{a} C_H \delta C_i \ ,  \\
    \delta C^{i}_j &= 0 \ \delta^i_j \ , 
\end{split}
\end{equation}
i.e. they cosmetically appear as pressureless matter violation ($w=0$). To make the analogy precise, however, one needs to check that the evolution itself is identical to that of matter. As we now show, this can be derived just by using the spatial components of Einstein's equations and covariant conservation of the visible matter.

By computing the variation of the Einstein's tensor to linear order in $\epsilon$ and matching terms order by order in an $\epsilon$ expansion, one obtains on top of the background Friedmann equations the linearised equations for the perturbations:
\begin{align}
    &4\pi G a^2 (\delta\rho + \delta C_H)= -k^2 \psi- 3\mathcal{H}(\psi'+\mathcal{H}\phi) \\
    &4\pi G a (1+w)\rho (\delta U_i+\delta C_i  = - i k_i (\psi' +\mathcal{H \phi}) \\
    &4\pi G a^2 w \delta\rho \ \delta^i_j  = \left[\psi'' + \mathcal{H}(2\psi + \phi)' + (2\mathcal{H}'+\mathcal{H}^2) \phi-\frac{1}{2}k^2 (\phi-\psi) \right] \delta^j_j - \frac{1}{2}(\phi-\psi)k_i k_j \ ,
\end{align}
where $k^2$ is the squared euclidean norm of $k$, $\mathcal{H}=a'/a$ and we indicate differentiation with respect of conformal time with a prime. The absence of anisotropic stress, as in standard cosmological perturbation theory, forces the two scalar metric functions to be equal to each other ($\psi=\phi$), reducing the equations of motion to:
\begin{align}
    &4\pi G a^2 (\delta\rho + \delta C_H) = -k^2 \phi- 3\mathcal{H}(\phi'+\mathcal{H}\phi) \label{eq:cons_inh} \\
    &4\pi G a \left[(1+w)\rho \delta U_i+ C_H \delta C_i \right]  = - i k_i (\phi' +\mathcal{H \phi})\label{eq:momcons_inh} \\
    &4\pi G a^2 w \delta\rho   = \left[\phi'' + \mathcal{H}(2\phi + \phi)' + (2\mathcal{H}'+\mathcal{H}^2) \phi \right] \ \label{eq:eom_inh} ,
\end{align}
from which we see that $\delta C$ acts as a matter perturbation and $\delta C_i$ as the velocity of the effective fluid. Since we minimally couple the noise of gravity, we can take the stress-tensor of visible matter to be covariantly conserved, which implies:
\begin{align}
\label{eq:continuity}
    &\delta \rho ' = -3 \mathcal{H}(1+w) \delta \rho - (1+w) \rho (i k_i \delta U_i - 3\phi') \\
\label{eq:euler}
    &\frac{1}{a^4}\left(a^5 (1+w) \rho  \ i k_i \delta U^i \right)'= w k^2 \delta \rho + (1+w) k^2 \rho \phi  .
\end{align}
It is now a matter of simple algebra to show that also $\delta C$ and $\delta C^i$, on shell, obey identical relations with $w=0$. It will be useful to use Friedmann equations, which correspond to:
\begin{equation}
\label{eq:friedmann_back}
    \mathcal{H}^2-\mathcal{H}' = 4\pi G a^2 \left[(1+w)\rho + C_H \right] \ .
\end{equation}

To derive the induced evolution equation for $\delta C$, we simply differentiate \eqref{eq:cons_inh} with respect to conformal time and substitute back \eqref{eq:momcons_inh}, \eqref{eq:eom_inh}, \eqref{eq:continuity} and \eqref{eq:friedmann_back} to obtain:
\begin{equation}
    \delta C' = -3 \mathcal{H}^2 \delta C - C_H \left(i k_i \delta C^i -3 \phi' \right) \ .
\end{equation}
This indeed matches \eqref{eq:continuity} with $w=0$. Similarly, taking both the time and spatial derivative of \eqref{eq:momcons_inh} and using the field equations of motion \eqref{eq:eom_inh} and \eqref{eq:euler} together with Friedmann's equation \eqref{eq:friedmann_back}, we obtain for the momentum constraint violation
\begin{equation}
    \frac{1}{a^4}\left(a^5 C_H  \ i k_i \delta C^i \right)'=  k^2 C_H \phi \ .
\end{equation}
Therefore, we see explicitly that, even at the inhomogeneous level, small constraint violations in the zero-noise limit behave exactly as pressureless dust perturbations with average energy density $C_H$. Indeed, the linearised inhomogeneous Hamiltonian constraint violation $\delta C$ acts as energy density perturbations, whilst the violation to the momentum constraint $\delta C^i$ plays the role of the peculiar velocity of the phantom fluid. 

\section{Stochastic constraint}\label{app:constraint}
We derive here the evolution of the constraint
\begin{equation}
    C_H= \frac{2 \pi G}{3} \frac{\pi_a^2}{a}-\rho a^3 + \frac{3 k a}{8\pi G}- \frac{a^3 \Lambda}{8\pi G}\ .
\end{equation}
on the stochastic trajectories given by
\begin{align}
    \dot{a} &= -\frac{4\pi G}{3} N \frac{\pi_a}{a} \\
    \dot{\pi}_a &= - \frac{2 \pi G}{3} N\frac{\pi_a^2}{a^2}+N 3 wa^2 \rho + N \frac{3}{8\pi G} k - N\frac{3}{8\pi G} \Lambda a^2+ \frac{3}{8\pi G}N a^2 \bar{\xi} \ , \\
    \rho &= \rho_0 a^{-3(1+w)} \ ,
\end{align}
i.e. with conservation of energy enforced in the matter sector. We allow:
\begin{equation}
    \mathbb{E}[\xi(t)]=0 \ , \qquad \mathbb{E}[\xi(t), \xi(t')]=\frac{D_2(a)}{N}\delta(t,t') \ ,
\end{equation}
i.e. the diffusion having a general dependence on the scale factor. It is always possible to rescale $\bar{\xi}$ as:
\begin{equation}
    \bar{\xi} = \sqrt{D_2(a,\pi_a)} \bar{\zeta} = \sqrt{D_2(a)} \frac{1}{\sqrt{N}} \frac{\text{d}W_t}{\text{d}t} \,
\end{equation}
such that we make the scaling of the noise with the fundamental degree of freedom $a$ manifest in the equations of motion (keeping, however, $\bar{\zeta}$ a scalar under time-parametrisation since we have not extracted the dependence on $N$). We later explore whether one can modify the equations of motion to preserve the value of the constraint at the level of the trajectories.

Using It\^{o}'s lemma, is then immediate to check that up to $\mathcal{O}(\text{d}t)$:
\begin{equation}
\begin{split}
\label{eq:constr_evo}
    \text{d}C_H &= \frac{\partial C_H}{\partial a} \text{d}a +  \frac{\partial C_H}{\partial \pi_a} \text{d} \pi_a + \frac{1}{2} \frac{\partial^2 C_H}{\partial \pi_a^2}  \text{d}\pi_a^2 \\
    &= \frac{1}{2}  a \pi_a \sqrt{N D_2(a,\pi_a)}\text{d}W_t+D_2(a,\pi_a) \frac{3}{32\pi G} a^3 N\text{d}t  \ ,
\end{split}
\end{equation}
which is indeed \eqref{eq:C-preservation} once the definition of the reparametrisation invariant white noise field is used and we declare $D_2(a,\pi_a) = 3 D_2 H^3/4\pi$. Only terms proportional to $D_2$ appear since the deterministic equations preserve the constraint, meaning that all the drift terms cancel each others in the algebra from the second to the third line of \eqref{eq:constr_evo}

As argued in the main body, the deviation from the deterministic constraint is not necessarily a sign that the stochastic theory is inconsistent. Still, a natural question is whether one can come up with any modification to the evolution process that conserves the deterministic constraint. It turns out that, for this simple system, there are two possible modifications that will lead to that result. The first one is coupling the noise to the matter system, breaking covariant conservation of energy-momentum for matter. We can see this with a bottom-up approach. Assume that the energy density is allowed to evolve via under a general SDE:
\begin{equation}
    \text{d}\rho = \left(4\pi G\rho \frac{\pi_a}{a^2}N+\mu(a,\pi_a,\rho) \right)\text{d}t + \sigma(a,\pi_a,\rho) \text{d}W_t \ ,
\end{equation}
where we have isolated the standard drift term. We can then \textit{require} that the constraint is satisfied and work out what the evolution law for $\rho$ needs to be. Using It\^{o}'s lemma once more, we get (note that $C_H$ is linear in $\rho$, so there is no spurious drift associated to its stocasticity):
\begin{equation}
\begin{split}
    \text{d}C_H &= \frac{\partial C_H}{\partial a} \text{d}a +  \frac{\partial C_H}{\partial \pi_a} \text{d} \pi_a + \frac{1}{2} \frac{\partial^2 C_H}{\partial \pi_a^2}  \text{d}\pi_a^2 + \frac{\partial C_H}{\partial \rho} \text{d}\rho \\
    &= \frac{1}{2}  a \pi_a \sqrt{N D_2}\text{d}W_t+\frac{3D_2 }{32\pi G} a^3 N\text{d}t  - a^3 \left(\mu(a,\pi_a,\rho) \text{d}t +\sigma(a,\pi_a,\rho) \text{d}W_t \right)   \\
    &= \left( \frac{1}{2}  a \pi_a \sqrt{N D_2}  - a^3 \sigma \right)\text{d}W_t + \left(\frac{3 D_2}{32\pi G}N a^3 -a^3 \mu\right) \text{d}t  \ .
\end{split}
\end{equation}
Requiring this to vanish amounts to setting:
\begin{align}
    \mu(a,\pi_a,\rho) &= \frac{3 D_2}{32\pi G}N  \\
    \sigma(a,\pi_a,\rho) &= \frac{\pi_a}{2a^2}   \sqrt{N D_2} \ .
\end{align}
Of course, for $D_2=0$ this recovers the deterministic evolution for matter.

\bibliography{PCDM}

@article{maartens2011is-the-universe,
	author = {Maartens, Roy},
	doi = {10.1098/rsta.2011.0289},
	eprint = {https://royalsocietypublishing.org/doi/pdf/10.1098/rsta.2011.0289},
	journal = {Philosophical Transactions of the Royal Society A: Mathematical, Physical and Engineering Sciences},
	number = {1957},
	pages = {5115-5137},
	title = {Is the Universe homogeneous?},
	url = {https://royalsocietypublishing.org/doi/abs/10.1098/rsta.2011.0289},
	volume = {369},
	year = {2011}}

@article{panella2024three-dimensional,
	archiveprefix = {arXiv},
	author = {Panella, Emanuele and Pedraza, Juan F. and Svesko, Andrew},
	eprint = {2407.03410},
	month = {7},
	primaryclass = {hep-th},
	reportnumber = {IFT-UAM/CSIC-24-98},
	title = {{Three-dimensional quantum black holes: a primer}},
	year = {2024}}

@article{climent2024chemical,
	archiveprefix = {arXiv},
	author = {Climent, Ana and Emparan, Roberto and Hennigar, Robie A.},
	eprint = {2404.15148},
	month = {4},
	primaryclass = {hep-th},
	title = {{Chemical Potential and Charge in Quantum Black Holes}},
	year = {2024}}

@misc{singh2015possible,
	archiveprefix = {arXiv},
	author = {Tejinder P. Singh},
	eprint = {1503.01040},
	primaryclass = {quant-ph},
	title = {Possible role of gravity in collapse of the wave-function: a brief survey of some ideas},
	url = {https://arxiv.org/abs/1503.01040},
	year = {2015}}

@article{goroff1985ultraviolet,
	author = {Goroff, Marc H. and Sagnotti, Augusto},
	doi = {10.1016/0550-3213(86)90193-8},
	journal = {Nucl. Phys. B},
	pages = {709--736},
	reportnumber = {CALT-68-1289, LBL-19995, UCB-PTH-85-34},
	title = {{The Ultraviolet Behavior of Einstein Gravity}},
	volume = {266},
	year = {1986}}

@article{julve1978quantum,
	author = {Julve, J. and Tonin, M.},
	doi = {10.1007/BF02748637},
	journal = {Nuovo Cim. B},
	pages = {137--152},
	reportnumber = {IFPD 2/78},
	title = {{Quantum Gravity with Higher Derivative Terms}},
	volume = {46},
	year = {1978}}

@article{fradkin1981renormalizable,
	author = {Fradkin, E. S. and Tseytlin, Arkady A.},
	doi = {10.1016/0370-2693(81)90702-4},
	journal = {Phys. Lett. B},
	pages = {377--381},
	title = {{Renormalizable Asymptotically Free Quantum Theory of Gravity}},
	volume = {104},
	year = {1981}}

@article{gisin1990weinbergs,
	author = {Gisin, Nicolas},
	journal = {Physics Letters A},
	number = {1-2},
	pages = {1--2},
	publisher = {Elsevier},
	title = {Weinberg's non-linear quantum mechanics and supraluminal communications},
	volume = {143},
	year = {1990}}

@article{polchinski1991weinbergs,
	author = {Polchinski, Joseph},
	journal = {Physical Review Letters},
	number = {4},
	pages = {397},
	publisher = {APS},
	title = {Weinberg's nonlinear quantum mechanics and the Einstein-Podolsky-Rosen paradox},
	volume = {66},
	year = {1991}}

@article{benedetti2009asymptotic,
	archiveprefix = {arXiv},
	author = {Benedetti, Dario and Machado, Pedro F. and Saueressig, Frank},
	doi = {10.1142/S0217732309031521},
	eprint = {0901.2984},
	journal = {Mod. Phys. Lett. A},
	pages = {2233--2241},
	primaryclass = {hep-th},
	reportnumber = {PI-QC-115, ITP-UU-09-03, SPIN-09-03, IPHT-T09-012},
	title = {{Asymptotic safety in higher-derivative gravity}},
	volume = {24},
	year = {2009}}

@article{akrami2020planck,
	author = {Akrami, Yashar and Arroja, Frederico and Ashdown, M and Aumont, J and Baccigalupi, Carlo and Ballardini, M and Banday, Anthony J and Barreiro, RB and Bartolo, Nicola and Basak, S and others},
	journal = {Astronomy \& Astrophysics},
	pages = {A10},
	publisher = {EDP sciences},
	title = {Planck 2018 results-X. Constraints on inflation},
	volume = {641},
	year = {2020}}

@article{oppenheim2023path,
	archiveprefix = {arXiv},
	author = {Jonathan Oppenheim and Zachary Weller-Davies},
	eprint = {2301.04677},
	primaryclass = {quant-ph},
	title = {Path integrals for classical-quantum dynamics},
	url = {https://arxiv.org/abs/2301.04677},
	year = {2023}}

@article{oppenheim2023covariant,
	archiveprefix = {arXiv},
	author = {Jonathan Oppenheim and Zachary Weller-Davies},
	eprint = {2302.07283},
	primaryclass = {gr-qc},
	title = {Covariant path integrals for quantum fields back-reacting on classical space-time},
	url = {https://arxiv.org/abs/2302.07283},
	year = {2023}}

@article{oppenheim2023is-it-time,
	author = {Oppenheim, Jonathan},
	doi = {10.1142/S0218271823420245},
	journal = {International Journal of Modern Physics D},
	publisher = {World Scientific},
	title = {Is it time to rethink quantum gravity?},
	year = {2023}}

@misc{kuchar1992time,
	author = {Kuchar, K},
	publisher = {Singapore: World Scientific},
	title = {Time and interpretation of quantum gravity Proc. 4th Canadian Conf. on General Relativity and Relativistic Astrophyscis},
	year = {1992}}

@article{oppenheim2009fundamental,
	archiveprefix = {arXiv},
	author = {Jonathan Oppenheim and Benni Reznik},
	eprint = {0902.2361},
	primaryclass = {hep-th},
	title = {Fundamental destruction of information and conservation laws},
	url = {https://arxiv.org/abs/0902.2361},
	year = {2009}}

@article{grudka2024renormalisation,
	archiveprefix = {arXiv},
	author = {Andrzej Grudka and Jonathan Oppenheim and Andrea Russo and Muhammad Sajjad},
	eprint = {2402.17844},
	primaryclass = {hep-th},
	title = {Renormalisation of postquantum-classical gravity},
	year = {2024}}

@article{buccio2024physical,
	archiveprefix = {arXiv},
	author = {Diego Buccio and John F. Donoghue and Gabriel Menezes and Roberto Percacci},
	eprint = {2403.02397},
	primaryclass = {hep-th},
	title = {Physical running of couplings in quadratic gravity},
	year = {2024}}

@unpublished{oppenheim2021constraints,
	author = {Jonathan Oppenheim},
	note = {manuscript in preparation},
	title = {The constraints of a continuous realisation of hybrid classical-quantum gravity}}

@article{anderson2012problem,
	archiveprefix = {arXiv},
	author = {Edward Anderson},
	eprint = {1009.2157},
	primaryclass = {gr-qc},
	title = {The Problem of Time in Quantum Gravity},
	url = {https://arxiv.org/abs/1009.2157},
	year = {2012}}

@article{isham1992canonical,
	archiveprefix = {arXiv},
	author = {Chris J. Isham},
	eprint = {gr-qc/9210011},
	primaryclass = {gr-qc},
	title = {Canonical Quantum Gravity and the Problem of Time},
	url = {https://arxiv.org/abs/gr-qc/9210011},
	year = {1992}}

@article{kuo1993semiclassical,
	author = {Kuo, Chung-I and Ford, L. H.},
	doi = {10.1103/physrevd.47.4510},
	issn = {0556-2821},
	journal = {Physical Review D},
	month = may,
	number = {10},
	pages = {4510--4519},
	publisher = {American Physical Society (APS)},
	title = {Semiclassical gravity theory and quantum fluctuations},
	url = {http://dx.doi.org/10.1103/PhysRevD.47.4510},
	volume = {47},
	year = {1993}}

@article{christensen1977trace,
	author = {Christensen, S. M. and Fulling, S. A.},
	doi = {10.1103/PhysRevD.15.2088},
	issue = {8},
	journal = {Phys. Rev. D},
	month = {Apr},
	numpages = {0},
	pages = {2088--2104},
	publisher = {American Physical Society},
	title = {Trace anomalies and the Hawking effect},
	url = {https://link.aps.org/doi/10.1103/PhysRevD.15.2088},
	volume = {15},
	year = {1977}}

@article{decesare2016effective,
	author = {Marco {de Cesare} and Fedele Lizzi and Mairi Sakellariadou},
	doi = {https://doi.org/10.1016/j.physletb.2016.07.015},
	issn = {0370-2693},
	journal = {Physics Letters B},
	pages = {498-501},
	title = {Effective cosmological constant induced by stochastic fluctuations of Newton's constant},
	url = {https://www.sciencedirect.com/science/article/pii/S0370269316303562},
	volume = {760},
	year = {2016}}

@article{arnowitt1959dynamical,
	author = {Arnowitt, R. and Deser, S. and Misner, C. W.},
	doi = {10.1103/PhysRev.116.1322},
	issue = {5},
	journal = {Phys. Rev.},
	month = {Dec},
	numpages = {0},
	pages = {1322--1330},
	publisher = {American Physical Society},
	title = {Dynamical Structure and Definition of Energy in General Relativity},
	url = {https://link.aps.org/doi/10.1103/PhysRev.116.1322},
	volume = {116},
	year = {1959}}

@article{layton2024classical,
	author = {Layton, Isaac and Oppenheim, Jonathan},
	doi = {10.1103/PRXQuantum.5.020331},
	issue = {2},
	journal = {PRX Quantum},
	month = {May},
	numpages = {27},
	pages = {020331},
	publisher = {American Physical Society},
	title = {The Classical-Quantum Limit},
	url = {https://link.aps.org/doi/10.1103/PRXQuantum.5.020331},
	volume = {5},
	year = {2024}}

@article{delgrosso2024cosmological,
	archiveprefix = {arXiv},
	author = {Del Grosso, Loris and Kaplan, David E. and Melia, Tom and Poulin, Vivian and Rajendran, Surjeet and Smith, Tristan L.},
	eprint = {2405.06374},
	month = {5},
	primaryclass = {hep-ph},
	title = {{Cosmological Consequences of Unconstrained Gravity and Electromagnetism}},
	year = {2024}}

@article{casadio2024relaxation,
	archiveprefix = {arXiv},
	author = {Casadio, Roberto and Chataignier, Leonardo and Kamenshchik, Alexander Yu. and Pedro, Francisco G. and Tronconi, Alessandro and Venturi, Giovanni},
	eprint = {2402.12437},
	month = {2},
	primaryclass = {gr-qc},
	title = {{Relaxation of first-class constraints and the quantization of gauge theories: from ''matter without matter'' to the reappearance of time in quantum gravity}},
	year = {2024}}

@article{burns2023time,
	archiveprefix = {arXiv},
	author = {Anne-Katherine Burns and David E. Kaplan and Tom Melia and Surjeet Rajendran},
	eprint = {2204.03043},
	primaryclass = {gr-qc},
	title = {Time Evolution in Quantum Cosmology},
	year = {2023}}

@article{moffat1997stochastic,
	author = {Moffat, J. W.},
	doi = {10.1103/physrevd.56.6264},
	issn = {1089-4918},
	journal = {Physical Review D},
	month = nov,
	number = {10},
	pages = {6264--6277},
	publisher = {American Physical Society (APS)},
	title = {Stochastic gravity},
	url = {http://dx.doi.org/10.1103/PhysRevD.56.6264},
	volume = {56},
	year = {1997}}

@article{wellerdavies2024quantum,
	archiveprefix = {arXiv},
	author = {Zachary Weller-Davies},
	eprint = {2402.17024},
	primaryclass = {gr-qc},
	title = {Quantum gravity with dynamical wave-function collapse via a classical scalar field},
	year = {2024}}

@article{risken1985fokker,
	adsurl = {https://ui.adsabs.harvard.edu/abs/1985JOSAB...2..508R},
	author = {{Risken}, H. and {Eberly}, J.~H.},
	journal = {Journal of the Optical Society of America B Optical Physics},
	month = mar,
	number = {3},
	pages = {508},
	title = {{The Fokker-Planck equation, methods of solution and applications}},
	volume = {2},
	year = 1985}

@article{alicki2003completely,
	author = {Alicki, Robert and Kryszewski, Stanis\l{}aw},
	doi = {10.1103/PhysRevA.68.013809},
	issue = {1},
	journal = {Phys. Rev. A},
	month = {Jul},
	numpages = {9},
	pages = {013809},
	publisher = {American Physical Society},
	title = {Completely positive Bloch-Boltzmann equations},
	url = {https://link.aps.org/doi/10.1103/PhysRevA.68.013809},
	volume = {68},
	year = {2003}}

@article{oppenheim2023gravitationally,
	author = {Oppenheim, Jonathan and Sparaciari, Carlo and {\v S}oda, Barbara and Weller-Davies, Zachary},
	date = {2023/12/04},
	doi = {10.1038/s41467-023-43348-2},
	id = {Oppenheim2023},
	isbn = {2041-1723},
	journal = {Nature Communications},
	number = {1},
	pages = {7910},
	title = {Gravitationally induced decoherence vs space-time diffusion: testing the quantum nature of gravity},
	url = {https://doi.org/10.1038/s41467-023-43348-2},
	volume = {14},
	year = {2023}}

@article{kaplan2023classical,
	archiveprefix = {arXiv},
	author = {Kaplan, David E. and Melia, Tom and Rajendran, Surjeet},
	eprint = {2305.01798},
	month = {5},
	primaryclass = {hep-th},
	reportnumber = {FERMILAB-PUB-23-227-SQMS-V},
	title = {{The Classical Equations of Motion of Quantized Gauge Theories, Part I: General Relativity}},
	year = {2023}}

@article{perez2021resolving,
	author = {Perez, Alejandro and Sudarsky, Daniel and Wilson-Ewing, Edward},
	doi = {10.1007/s10714-020-02781-0},
	issn = {1572-9532},
	journal = {General Relativity and Gravitation},
	month = jan,
	number = {1},
	publisher = {Springer Science and Business Media LLC},
	title = {Resolving the H0 tension with diffusion},
	url = {http://dx.doi.org/10.1007/s10714-020-02781-0},
	volume = {53},
	year = {2021}}

@article{landau2022cosmological,
	archiveprefix = {arXiv},
	author = {Susana J. Landau and Micol Benetti and Alejandro Perez and Daniel Sudarsky},
	eprint = {2211.07424},
	primaryclass = {astro-ph.CO},
	title = {Cosmological constraints on unimodular gravity models with diffusion},
	year = {2022}}

@article{panella2023quantum,
	author = {Panella, Emanuele and Svesko, Andrew},
	doi = {10.1007/jhep06(2023)127},
	issn = {1029-8479},
	journal = {Journal of High Energy Physics},
	month = jun,
	number = {6},
	publisher = {Springer Science and Business Media LLC},
	title = {Quantum Kerr-de Sitter black holes in three dimensions},
	url = {http://dx.doi.org/10.1007/JHEP06(2023)127},
	volume = {2023},
	year = {2023}}

@article{emparan2022black,
	author = {Emparan, Roberto and Pedraza, Juan F. and Svesko, Andrew and Toma{\v s}evi{\'c}, Marija and Visser, Manus R.},
	doi = {10.1007/jhep11(2022)073},
	issn = {1029-8479},
	journal = {Journal of High Energy Physics},
	month = nov,
	number = {11},
	publisher = {Springer Science and Business Media LLC},
	title = {Black holes in dS3},
	url = {http://dx.doi.org/10.1007/JHEP11(2022)073},
	volume = {2022},
	year = {2022}}

@article{emparan2020quantum,
	author = {Emparan, Roberto and Frassino, Antonia Micol and Way, Benson},
	doi = {10.1007/jhep11(2020)137},
	issn = {1029-8479},
	journal = {Journal of High Energy Physics},
	month = nov,
	number = {11},
	publisher = {Springer Science and Business Media LLC},
	title = {Quantum BTZ black hole},
	url = {http://dx.doi.org/10.1007/JHEP11(2020)137},
	volume = {2020},
	year = {2020}}

@article{emparan2002quantum,
	author = {Emparan, Roberto and Fabbri, Alessandro and Kaloper, Nemanja},
	doi = {10.1088/1126-6708/2002/08/043},
	issn = {1029-8479},
	journal = {Journal of High Energy Physics},
	month = aug,
	number = {08},
	pages = {043--043},
	publisher = {Springer Science and Business Media LLC},
	title = {Quantum Black Holes as Holograms in AdS Braneworlds},
	url = {http://dx.doi.org/10.1088/1126-6708/2002/08/043},
	volume = {2002},
	year = {2002}}

@article{oppenheim2024anomalous,
	archiveprefix = {arXiv},
	author = {Jonathan Oppenheim and Andrea Russo},
	eprint = {2402.19459},
	primaryclass = {gr-qc},
	title = {Anomalous contribution to galactic rotation curves due to stochastic spacetime},
	year = {2024}}

@book{dirac1967lectures,
	address = {New York},
	author = {Dirac, Paul A. M.},
	note = {Belfer Graduate School of Science Monographs Series},
	publisher = {Yeshiva University Press},
	title = {Lectures on Quantum Mechanics},
	year = {1967}}

@article{wands2000approach,
	author = {David Wands and Karim A. Malik and David H. Lyth and Andrew R. Liddle},
	journal = {Physical Review D},
	pages = {043527},
	title = {A New approach to the evolution of cosmological perturbations on large scales},
	url = {https://api.semanticscholar.org/CorpusID:40791120},
	volume = {62},
	year = {2000}}

@article{oppenheim2021postquantum,
	author = {Oppenheim, Jonathan},
	doi = {10.1103/PhysRevX.13.041040},
	issue = {4},
	journal = {Phys. Rev. X},
	month = {Dec},
	numpages = {37},
	pages = {041040},
	publisher = {American Physical Society},
	title = {A Postquantum Theory of Classical Gravity?},
	url = {https://link.aps.org/doi/10.1103/PhysRevX.13.041040},
	volume = {13},
	year = {2023}}

@article{hojman1976geometrodynamics,
	author = {Hojman, Sergio A and Kucha{\v r}, Karel and Teitelboim, Claudio},
	date = {1976/01/01/},
	doi = {https://doi.org/10.1016/0003-4916(76)90112-3},
	isbn = {0003-4916},
	journal = {Annals of Physics},
	number = {1},
	pages = {88--135},
	title = {Geometrodynamics regained},
	url = {https://www.sciencedirect.com/science/article/pii/0003491676901123},
	volume = {96},
	year = {1976}}

@article{kampen1981ito,
	author = {van Kampen, N. G.},
	date = {1981/01/01},
	doi = {10.1007/BF01007642},
	id = {van Kampen1981},
	isbn = {1572-9613},
	journal = {Journal of Statistical Physics},
	number = {1},
	pages = {175--187},
	title = {It{\^o} versus Stratonovich},
	url = {https://doi.org/10.1007/BF01007642},
	volume = {24},
	year = {1981}}

@article{oppenheim2024diffeomorphism,
	archiveprefix = {arXiv},
	author = {Oppenheim, Jonathan and Russo, Andrea and Weller-Davies, Zachary},
	doi = {10.1103/PhysRevD.110.024007},
	eprint = {2401.05514},
	journal = {Phys. Rev. D},
	number = {2},
	pages = {024007},
	primaryclass = {gr-qc},
	title = {{Diffeomorphism invariant classical-quantum path integrals for Nordstr\"om gravity}},
	volume = {110},
	year = {2024}}

@article{tilloy2017principle,
	author = {Tilloy, Antoine and Di\'osi, Lajos},
	doi = {10.1103/PhysRevD.96.104045},
	issue = {10},
	journal = {Phys. Rev. D},
	month = {Nov},
	numpages = {6},
	pages = {104045},
	publisher = {American Physical Society},
	title = {Principle of least decoherence for Newtonian semiclassical gravity},
	url = {https://link.aps.org/doi/10.1103/PhysRevD.96.104045},
	volume = {96},
	year = {2017}}

@article{tilloy2016sourcing,
	author = {Tilloy, Antoine and Di\'osi, Lajos},
	doi = {10.1103/PhysRevD.93.024026},
	issue = {2},
	journal = {Phys. Rev. D},
	month = {Jan},
	numpages = {12},
	pages = {024026},
	publisher = {American Physical Society},
	title = {Sourcing semiclassical gravity from spontaneously localized quantum matter},
	url = {https://link.aps.org/doi/10.1103/PhysRevD.93.024026},
	volume = {93},
	year = {2016}}

@article{kafri2014classical,
	author = {Kafri, D and Taylor, J M and Milburn, G J},
	date = {2014/06/26},
	doi = {10.1088/1367-2630/16/6/065020},
	isbn = {1367-2630;},
	journal = {New Journal of Physics},
	number = {6},
	pages = {065020},
	publisher = {IOP Publishing},
	title = {A classical channel model for gravitational decoherence},
	url = {https://dx.doi.org/10.1088/1367-2630/16/6/065020},
	volume = {16},
	year = {2014}}

@article{diosi2014hybrid,
	author = {Di{\'o}si, Lajos},
	date = {2014/12/19},
	doi = {10.1088/0031-8949/2014/T163/014004},
	isbn = {1402-4896; 0031-8949},
	journal = {Physica Scripta},
	number = {T163},
	pages = {014004},
	publisher = {IOP Publishing},
	title = {Hybrid quantum-classical master equations},
	url = {https://dx.doi.org/10.1088/0031-8949/2014/T163/014004},
	volume = {2014},
	year = {2014}}

@article{diosi1995quantum,
	archiveprefix = {arXiv},
	author = {Lajos Diosi},
	eprint = {quant-ph/9503023},
	primaryclass = {quant-ph},
	title = {Quantum dynamics with two Planck constants and the semiclassical limit},
	url = {https://arxiv.org/abs/quant-ph/9503023},
	year = {1995}}

@article{blanchard1995event-enhanced,
	author = {Blanchard, Ph. and Jadczyk, A.},
	doi = {https://doi.org/10.1002/andp.19955070605},
	journal = {Annalen der Physik},
	number = {6},
	pages = {583-599},
	title = {Event-enhanced quantum theory and piecewise deterministic dynamics},
	url = {https://onlinelibrary.wiley.com/doi/abs/10.1002/andp.19955070605},
	volume = {507},
	year = {1995}}

@article{ahmed2024semiclassical,
	author = {Ahmed, Shahnewaz and Lima, Caroline and Mart{\'\i}n-Mart{\'\i}nez, Eduardo},
	date = {2024/01/02},
	doi = {10.1007/JHEP01(2024)001},
	id = {Ahmed2024},
	isbn = {1029-8479},
	journal = {Journal of High Energy Physics},
	number = {1},
	pages = {1},
	title = {Semiclassical gravity beyond coherent states},
	url = {https://doi.org/10.1007/JHEP01(2024)001},
	volume = {2024},
	year = {2024}}

@article{oppenheim2022constraints,
	author = {Oppenheim, Jonathan and Weller-Davies, Zachary},
	date = {2022/02/10},
	doi = {10.1007/JHEP02(2022)080},
	id = {Oppenheim2022},
	isbn = {1029-8479},
	journal = {Journal of High Energy Physics},
	number = {2},
	pages = {80},
	title = {The constraints of post-quantum classical gravity},
	url = {https://doi.org/10.1007/JHEP02(2022)080},
	volume = {2022},
	year = {2022}}

@inbook{Oksendal:2023aa,
	address = {Cham %@ 978-3-031-12244-6},
	author = {{\O}ksendal, Bernt},
	booktitle = {Mathematics Going Forward : Collected Mathematical Brushstrokes},
	caption = {{\O}ksendal2023},
	date = {2023},
	pages = {629-649},
	publisher = {Springer International Publishing},
	title = {Space-Time Stochastic Calculus and White Noise},
	type = {10.1007/978-3-031-12244-6_44},
	url = {https://doi.org/10.1007/978-3-031-12244-6_44},
	year = {2023}}

@article{galley2023consistent,
	author = {Galley, Thomas D. and Giacomini, Flaminia and Selby, John H.},
	doi = {10.22331/q-2023-10-16-1142},
	issn = {2521-327X},
	journal = {{Quantum}},
	month = oct,
	pages = {1142},
	publisher = {{Verein zur F{\"{o}}rderung des Open Access Publizierens in den Quantenwissenschaften}},
	title = {Any consistent coupling between classical gravity and quantum matter is fundamentally irreversible},
	url = {https://doi.org/10.22331/q-2023-10-16-1142},
	volume = {7},
	year = {2023}}

@article{ahmed2004everpresent,
	author = {Ahmed, Maqbool and Dodelson, Scott and Greene, Patrick B. and Sorkin, Rafael},
	doi = {10.1103/physrevd.69.103523},
	issn = {1550-2368},
	journal = {Physical Review D},
	month = may,
	number = {10},
	publisher = {American Physical Society (APS)},
	title = {Everpresent Lambda},
	url = {http://dx.doi.org/10.1103/PhysRevD.69.103523},
	volume = {69},
	year = {2004}}

@article{misawa1999conserved,
	author = {Misawa, Tetsuya},
	date = {1999/12/01},
	doi = {10.1023/A:1004095516648},
	id = {Misawa1999},
	isbn = {1572-9052},
	journal = {Annals of the Institute of Statistical Mathematics},
	number = {4},
	pages = {779--802},
	title = {Conserved Quantities and Symmetries Related to Stochastic Dynamical Systems},
	url = {https://doi.org/10.1023/A:1004095516648},
	volume = {51},
	year = {1999}}

@article{albeverio1995remark,
	author = {S Albeverio and Shao-Ming Fei},
	date = {1995/11/21},
	doi = {10.1088/0305-4470/28/22/012},
	isbn = {0305-4470},
	journal = {Journal of Physics A: Mathematical and General},
	number = {22},
	pages = {6363},
	title = {A remark on symmetry of stochastic dynamical systems and their conserved quantities},
	url = {https://dx.doi.org/10.1088/0305-4470/28/22/012},
	volume = {28},
	year = {1995}}

@article{brown1995dust,
	author = {Brown, J. David and Kucha{\v r}, Karel V.},
	doi = {10.1103/physrevd.51.5600},
	issn = {0556-2821},
	journal = {Physical Review D},
	month = may,
	number = {10},
	pages = {5600--5629},
	publisher = {American Physical Society (APS)},
	title = {Dust as a standard of space and time in canonical quantum gravity},
	url = {http://dx.doi.org/10.1103/physrevd.51.5600},
	volume = {51},
	year = {1995}}

@article{layton2023weak,
	author = {Layton, Isaac and Oppenheim, Jonathan and Russo, Andrea and Weller-Davies, Zachary},
	date = {2023/08/24},
	doi = {10.1007/JHEP08(2023)163},
	id = {Layton2023},
	isbn = {1029-8479},
	journal = {Journal of High Energy Physics},
	number = {8},
	pages = {163},
	title = {The weak field limit of quantum matter back-reacting on classical spacetime},
	url = {https://doi.org/10.1007/JHEP08(2023)163},
	volume = {2023},
	year = {2023}}

@article{bergmann1957summary,
	author = {Bergmann, Peter G.},
	doi = {10.1103/RevModPhys.29.352},
	issue = {3},
	journal = {Rev. Mod. Phys.},
	month = {Jul},
	numpages = {0},
	pages = {352--354},
	publisher = {American Physical Society},
	title = {Summary of the Chapel Hill Conference},
	url = {https://link.aps.org/doi/10.1103/RevModPhys.29.352},
	volume = {29},
	year = {1957}}

@article{eppley1977tnecessity,
	author = {Eppley, Kenneth and Hannah, Eric},
	date = {1977/02/01},
	doi = {10.1007/BF00715241},
	id = {Eppley1977},
	isbn = {1572-9516},
	journal = {Foundations of Physics},
	number = {1},
	pages = {51--68},
	title = {The necessity of quantizing the gravitational field},
	url = {https://doi.org/10.1007/BF00715241},
	volume = {7},
	year = {1977}}

@article{aghanim2020planck,
	author = {N. Aghanim and Y. Akrami and M. Ashdown and J. Aumont and C. Baccigalupi and M. Ballardini and A. J. Banday and R. B. Barreiro and N. Bartolo and S. Basak and R. Battye and K. Benabed and J.-P. Bernard and M. Bersanelli and P. Bielewicz and J. J. Bock and J. R. Bond and J. Borrill and F. R. Bouchet and F. Boulanger and M. Bucher and C. Burigana and R. C. Butler and E. Calabrese and J.-F. Cardoso and J. Carron and A. Challinor and H. C. Chiang and J. Chluba and L. P. L. Colombo and C. Combet and D. Contreras and B. P. Crill and F. Cuttaia and P. de Bernardis and G. de Zotti and J. Delabrouille and J.-M. Delouis and E. Di Valentino and J. M. Diego and O. Dor{\'{e} } and M. Douspis and A. Ducout and X. Dupac and S. Dusini and G. Efstathiou and F. Elsner and T. A. En{\ss}lin and H. K. Eriksen and Y. Fantaye and M. Farhang and J. Fergusson and R. Fernandez-Cobos and F. Finelli and F. Forastieri and M. Frailis and A. A. Fraisse and E. Franceschi and A. Frolov and S. Galeotta and S. Galli and K. Ganga and R. T. G{\'{e}}nova-Santos and M. Gerbino and T. Ghosh and J. Gonz{\'{a}}lez-Nuevo and K. M. G{\'{o}}rski and S. Gratton and A. Gruppuso and J. E. Gudmundsson and J. Hamann and W. Handley and F. K. Hansen and D. Herranz and S. R. Hildebrandt and E. Hivon and Z. Huang and A. H. Jaffe and W. C. Jones and A. Karakci and E. Keih{\"a}nen and R. Keskitalo and K. Kiiveri and J. Kim and T. S. Kisner and L. Knox and N. Krachmalnicoff and M. Kunz and H. Kurki-Suonio and G. Lagache and J.-M. Lamarre and A. Lasenby and M. Lattanzi and C. R. Lawrence and M. Le Jeune and P. Lemos and J. Lesgourgues and F. Levrier and A. Lewis and M. Liguori and P. B. Lilje and M. Lilley and V. Lindholm and M. L{\'{o}}pez-Caniego and P. M. Lubin and Y.-Z. Ma and J. F. Mac{\'{\i}}as-P{\'{e}}rez and G. Maggio and D. Maino and N. Mandolesi and A. Mangilli and A. Marcos-Caballero and M. Maris and P. G. Martin and M. Martinelli and E. Mart{\'{\i}}nez-Gonz{\'{a}}lez and S. Matarrese and N. Mauri and J. D. McEwen and P. R. Meinhold and A. Melchiorri and A. Mennella and M. Migliaccio and M. Millea and S. Mitra and M.-A. Miville-Desch{\^{e}}nes and D. Molinari and L. Montier and G. Morgante and A. Moss and P. Natoli and H. U. N{\o}rgaard-Nielsen and L. Pagano and D. Paoletti and B. Partridge and G. Patanchon and H. V. Peiris and F. Perrotta and V. Pettorino and F. Piacentini and L. Polastri and G. Polenta and J.-L. Puget and J. P. Rachen and M. Reinecke and M. Remazeilles and A. Renzi and G. Rocha and C. Rosset and G. Roudier and J. A. Rubi{\~{n}}o-Mart{\'{\i}}n and B. Ruiz-Granados and L. Salvati and M. Sandri and M. Savelainen and D. Scott and E. P. S. Shellard and C. Sirignano and G. Sirri and L. D. Spencer and R. Sunyaev and A.-S. Suur-Uski and J. A. Tauber and D. Tavagnacco and M. Tenti and L. Toffolatti and M. Tomasi and T. Trombetti and L. Valenziano and J. Valiviita and B. Van Tent and L. Vibert and P. Vielva and F. Villa and N. Vittorio and B. D. Wandelt and I. K. Wehus and M. White and S. D. M. White and A. Zacchei and A. Zonca},
	doi = {10.1051/0004-6361/201833910},
	journal = {Astronomy {\&} Astrophysics},
	month = {sep},
	pages = {A6},
	publisher = {{EDP} Sciences},
	title = {Planck 2018 results - VI. Cosmological parameters},
	url = {https://doi.org/10.1051%2F0004-6361%2F201833910},
	volume = {641},
	year = 2020}

@article{oppenheim2022classes,
	archiveprefix = {arXiv},
	author = {Jonathan Oppenheim and Carlo Sparaciari and Barbara {\v S}oda and Zachary Weller-Davies},
	eprint = {2203.01332},
	primaryclass = {quant-ph},
	title = {The two classes of hybrid classical-quantum dynamics},
	year = {2022}}

@article{layton2023healthier,
	archiveprefix = {arXiv},
	author = {Isaac Layton and Jonathan Oppenheim and Zachary Weller-Davies},
	eprint = {2208.11722},
	primaryclass = {quant-ph},
	title = {A healthier semi-classical dynamics},
	year = {2023}}

@article{hu2008stochastic,
	author = {Hu, Bei Lok and Verdaguer, Enric},
	date = {2008/12/01},
	doi = {10.12942/lrr-2008-3},
	id = {Hu2008},
	isbn = {1433-8351},
	journal = {Living Reviews in Relativity},
	number = {1},
	pages = {3},
	title = {Stochastic Gravity: Theory and Applications},
	url = {https://doi.org/10.12942/lrr-2008-3},
	volume = {11},
	year = {2008}}

@article{brandenberger2016bouncing,
    author = "Brandenberger, Robert and Peter, Patrick",
    title = "{Bouncing Cosmologies: Progress and Problems}",
    eprint = "1603.05834",
    archivePrefix = "arXiv",
    primaryClass = "hep-th",
    doi = "10.1007/s10701-016-0057-0",
    journal = "Found. Phys.",
    volume = "47",
    number = "6",
    pages = "797--850",
    year = "2017"
}

@article{ashtekar2006quantum,
  title = {Quantum Nature of the Big Bang},
  author = {Ashtekar, Abhay and Pawlowski, Tomasz and Singh, Parampreet},
  journal = {Phys. Rev. Lett.},
  volume = {96},
  issue = {14},
  pages = {141301},
  numpages = {4},
  year = {2006},
  month = {Apr},
  publisher = {American Physical Society},
  doi = {10.1103/PhysRevLett.96.141301},
  url = {https://link.aps.org/doi/10.1103/PhysRevLett.96.141301}
}

@article{battefeld2014critical,
    author = "Battefeld, D. and Peter, Patrick",
    title = "{A Critical Review of Classical Bouncing Cosmologies}",
    eprint = "1406.2790",
    archivePrefix = "arXiv",
    primaryClass = "astro-ph.CO",
    doi = "10.1016/j.physrep.2014.12.004",
    journal = "Phys. Rept.",
    volume = "571",
    pages = "1--66",
    year = "2015"
}

@article{friedmann1922curvature,
  author  = {Friedmann, Alexander A.},
  title   = {\"{U}ber die Kr\"{u}mmung des Raumes},
  journal = {Zeitschrift f{\"u}r Physik},
  volume  = {10},
  pages   = {377--386},
  year    = {1922},
  doi     = {10.1007/BF01332580},
  url     = {https://doi.org/10.1007/BF01332580}
}

@article{tolman1931periodic,
  author  = {Tolman, Richard C.},
  title   = {On the Theoretical Requirements for a Periodic Behaviour of the Universe},
  journal = {Physical Review},
  volume  = {38},
  pages   = {1758--1771},
  year    = {1931},
  doi     = {10.1103/PhysRev.38.1758},
  url     = {https://doi.org/10.1103/PhysRev.38.1758}
}

@article{oppenheim2025diffusion,
    author = "Oppenheim, Jonathan and Panella, Emanuele",
    title = "{Diffusion in the stochastic Klein-Gordon equation}",
    eprint = "2511.10738",
    archivePrefix = "arXiv",
    primaryClass = "gr-qc",
    month = "11",
    year = "2025"
}

@article{bishop2019onedimensional,
   title={On one-dimensional Riccati diffusions},
   volume={29},
   ISSN={1050-5164},
   url={http://dx.doi.org/10.1214/18-AAP1431},
   DOI={10.1214/18-aap1431},
   number={2},
   journal={The Annals of Applied Probability},
   publisher={Institute of Mathematical Statistics},
   author={Bishop, A. N. and Del Moral, P. and Kamatani, K. and Rémillard, B.},
   year={2019},
   month=apr }

\end{document}